\pdfoutput=1
\documentclass[12pt]{article}
\usepackage[bookmarks=true]{hyperref}
\usepackage{graphicx,epsfig,color}
\usepackage{cite}
\usepackage[english]{babel}
\usepackage{amssymb,amsfonts,amsmath}
\usepackage{verbatim}
\usepackage{physics}
\usepackage{cancel}
\usepackage[top=2.5cm, bottom=2.5cm, left=2cm, right=2cm]{geometry}	
\usepackage{tikz}
\usetikzlibrary{shapes}

\newcommand{\be}{\begin{equation}}
\newcommand{\ee}{\end{equation}}
\newcommand{\bea}{\begin{eqnarray}}
\newcommand{\eea}{\end{eqnarray}}

\def\ie{{\em i.e.},}
\def\eg{{\em e.g.},}

\newcommand{\mt}[1]{\textrm{\scriptsize #1}}

\def\EW{E_\mt{W}}
\def\EP{E_\mt{P}}

\graphicspath{{figures/}}
\numberwithin{equation}{section}

\begin{document}

\begin{flushright}
APCTP Pre2022 - 24 \\
HIP-2022-27/TH
\end{flushright}

\begin{center}

\centerline{\Large {\bf  Bounding entanglement wedge cross sections}}

\vspace{8mm}

\renewcommand\thefootnote{\mbox{$\fnsymbol{footnote}$}}
Parul Jain,${}^{1}$\footnote{parul.jain@apctp.org} 
Niko Jokela,${}^{2,3}$\footnote{niko.jokela@helsinki.fi} Matti J\"arvinen,${}^{1,4}$\footnote{matti.jarvinen@apctp.org} and Subhash Mahapatra${}^5$\footnote{mahapatrasub@nitrkl.ac.in}

\vspace{4mm}

${}^1${\small \sl Asia Pacific Center for Theoretical Physics} \\
{\small \sl Pohang 37673, Republic of Korea} 
\vskip 0.2cm

${}^2${\small \sl Department of Physics} and ${}^3${\small \sl Helsinki Institute of Physics} \\
{\small \sl P.O.Box 64} \\
{\small \sl FIN-00014 University of Helsinki, Finland} 
\vskip 0.2cm

${}^4${\small \sl Department of Physics, Pohang University of Science and Technology} \\
{\small \sl Pohang 37673, Republic of Korea} 
\vskip 0.2cm

${}^5${\small \sl Department of Physics and Astronomy} \\
{\small \sl National Institute of Technology Rourkela} \\
{\small \sl Rourkela - 769008, India} 
\vskip 0.2cm

\end{center}

\vspace{6mm}
\numberwithin{equation}{section}
\setcounter{footnote}{0}
\renewcommand\thefootnote{\mbox{\arabic{footnote}}}

\begin{abstract} 
The entanglement wedge cross sections (EWCSs) are postulated as dual gravity probes to certain measures for the entanglement of multiparty systems.  We test various proposed inequalities for EWCSs. As it turns out, contrary to expectations, the EWCS is not clearly monogamous nor polygamous for tripartite systems but the results depend on the details and dimensionality of the geometry of the gravity solutions. We propose weaker monogamy relations for dual entanglement measures, which lead to a new lower bound on EWCS. Our work is based on a plethora of gravity backgrounds: pure anti de Sitter spaces, anti de Sitter black branes, those induced by a stack of D$p$-branes, and cigar geometries in generic dimension. 
\end{abstract}

\newpage
\tableofcontents

\newpage

\section{Introduction}\label{sec:intro}

Remarkable connections between seemingly different fields of physics, namely quantum information and quantum gravity, have been realized in recent years. At the heart of these connections lie entanglement measures that have been explored in the context of AdS/CFT correspondence. In particular, the Ryu-Takayanagi (RT) proposal \cite{Ryu:2006ef,Ryu:2006bv} relates an entanglement entropy (EE) associated with a subregion in the boundary field theory to that of a minimal bulk surface in the dual gravity setup. This proposal has prompted considerable amount of studies, in particular, since it has the advantage of enabling the reconstruction of various aspects of the gravity theory purely from the entanglement data at the boundary \cite{Almheiri:2014lwa,Dong:2016eik,Bao:2015bfa,Jokela:2020auu}. The holographic EE has passed several non-trivial tests such as satisfying the strong subadditivity and others that quantum states are expected to obey \cite{Ryu:2006ef,Headrick:2013zda,Hayden:2011ag,Headrick:2007km,Bao:2015bfa}. The RT conjecture therefore may place constraints on viable classical holographic spacetimes \cite{Hayden:2011ag}.

It is interesting to ask whether there exist holographic definitions of other entanglement measures and, if yes, what type of inequalities and constraints these entanglement measures should obey. One particular bulk object that has attracted a great deal of interest in recent years is the entanglement wedge cross section (EWCS) $\EW$, the minimal area of the surface that bipartitions the entanglement wedge region \cite{Czech:2012bh,Wall:2012uf,Headrick:2014cta}. $\EW$ was originally proposed in \cite{Takayanagi:2017knl} to capture the entanglement of purification $\EP$ for bipartite mixed states. The dual interpretation is still open, however, many other dual candidate information theoretic quantities for $\EW$ have emerged.  For instance, $\EW$ has been suggested to act as the holographic dual to odd EE \cite{Tamaoka:2018ned} as well as to the reflected EE \cite{Dutta:2019gen}. Similarly, it also appears in the holographic proposal of the entanglement negativity \cite{Kudler-Flam:2018qjo,Kusuki:2019zsp}. This development is particularly interesting since, unlike the EE, $\EW$ is better fit to measure the entanglement of mixed states. This is also important from the application point of view as only a handful of examples are known in the field theory side where these mixed state entanglement measures can be computed explicitly. Though the issue concerning the correct boundary interpretation of $\EW$ is still far from being fully settled, however, a great deal of work has been done in exploring its properties in various holographic settings, see \cite{Bhattacharyya:2018sbw,Hirai:2018jwy,Espindola:2018ozt,Bao:2018gck,Umemoto:2018jpc,Yang:2018gfq,Caputa:2018xuf,Liu:2019qje,Bhattacharyya:2019tsi,BabaeiVelni:2019pkw,Jokela:2019ebz,Bao:2019wcf,Harper:2019lff,Jeong:2019xdr,Bao:2019zqc,Chu:2019etd,Akers:2019gcv,Du:2019vwh,Kudler-Flam:2020url,Chandrasekaran:2020qtn,Li:2020ceg,Lala:2020lcp,Jain:2020rbb,Jokela:2020wgs,Wen:2021qgx,Bao:2021vyq,Basu:2021awn,Hayden:2021gno,Akers:2021pvd,Camargo:2022mme,Jain:2022hxl} for progress in this vein.  

Another interesting direction is to explore multipartite correlation measures and their inequality relations. In quantum systems, the entanglement structure and separability criteria of states are much richer in multipartite case compared to bipartite and therefore are interesting to analyze. For instance, multipartite measures put much stricter constraints on entanglement sharing. Indeed,  one of the most important features of multipartite quantum systems is that entanglement is monogamous \cite{Horodecki:2009zz}. It is a statement that if two systems are strongly entangled then they can not be strongly entangled with another system at the same time. The monogamy, for some measure of entanglement $f$, is generally stated as an inequality
\begin{eqnarray}
 f(A:BC) \geq f(A:B) + f(A:C) \ , \label{monogamycondition}
\end{eqnarray}
which mathematically emulates the fact that entangled correlations between $A$ and $B$ limit their entanglement with $C$. Some entanglement measures where such monogamy relations have been explicitly verified in quantum systems are squared concurrence and entanglement negativity \cite{PhysRevA.61.052306, PhysRevA.75.062308, PhysRevA.91.012339}. Since this trade-off between the amount of entanglement between systems is purely quantum in nature with no classical analogue, it therefore provides a way to characterize the structure of multipartite entanglement. Indeed, this has been the subject of intense discussion in the quantum information community in recent years \cite{Horodecki:2009zz}. Let us also emphasize that monogamy is not an intrinsic property of other quantum correlation measures. Indeed, there are entanglement measures, such as the relative entropy of entanglement or the entanglement of formation, that do not satisfy monogamy criterion (\ref{monogamycondition}) in general.

In the context of holography much less is known about the multipartite entanglement and its properties compared to the bipartite configuration. The holographic tripartite information, which is a generalization of the mutual information, has been shown to be monogamous \cite{Hayden:2011ag}. The generalization to $n$-partite information further shows that the sign of $n$-partite information constrains the monogamy of $(n-1)$-partite information \cite{Alishahiha:2014jxa,Mirabi:2016elb}. Interestingly, the EWCS, on the other hand, has been proposed to be polygamous for a tripartite pure state \cite{Takayanagi:2017knl,Nguyen:2017yqw}, meaning 
\be\label{polygamycondition}
 \EW(A:B) + \EW(A:C) \geq \EW(A:BC)  \ .
\ee
Further, in \cite{Takayanagi:2017knl}, the following property for the $\EW$ was also proposed
\be\label{TakaIneq}
\EW(A:BC) \geq \EW(A:B) \ .
\ee
In addition, $\EW$ satisfies 
\be\label{EWineqality}
 {\rm{min}}(S_A,S_B) \geq \EW(A:B) \geq \frac{1}{2} I(A:B) \ .
\ee
It is important to note here that the geometric proof exists only for the inequality (\ref{EWineqality}).

These inequalities of $\EW$ were the reasons which lead to the holographic proposal of it being dual to the entanglement of purification, as it seemed to obey the same set of inequalities that the entanglement of purification is known to obey. These and other inequalities for the entanglement of purification can be found in \cite{PhysRevA.91.042323}.  Analogous inequalities for $\EW$ have not been explicitly tested; for work in this direction, see \cite{Umemoto:2018jpc}. Besides the systematic scan of whether inequalities hold, we already know other some peculiarities of $\EW$, namely, that $\EW$ is neither continuous nor monotonic under a renormalization group flow in the large number of colors $N$ limit \cite{Jokela:2019ebz}. 

The extensive study of inequalities for $\EW$ is important as it not only sheds light on the unsettled status of the holographic interpretation of $\EW$, but will allow us to constrain other multipartite measures, such as the multipartite mutual information \cite{Akers:2019gcv}. Alternatively, if $\EW$ is to satisfy superadditivity, which does not hold for a generic state given a boundary quantum field theory, will lead to a way to discern which quantum states are to have holographic duals. 

In this work, we put the suggested inequalities for $\EW$ to test in various holographic settings. We consider pure AdS, AdS black branes, AdS soliton (``cigar" shaped), and non-conformal D$p$-brane geometries in various dimensions and analyze the validity of above-mentioned inequalities for $\EW$. In all of the cases we study in this article we found that the inequalities (\ref{TakaIneq}) and (\ref{EWineqality}) are always satisfied. Interestingly, however, in the parameter space of $l$ and $s$, where $l$ is the slab width and $s$ is the separation, there are regions where the polygamy of $\EW$ can be violated. In particular, Eq.~(\ref{polygamycondition}) is violated for all configurations in pure AdS$_3$. We have checked this result both analytically and numerically. Similarly,  we find evidence for the violation of (\ref{polygamycondition}) for all values of $d$ in case of a generic $(d+1)$-dimensional pure AdS, AdS black brane, and AdS soliton geometries. The polygamy condition is in addition violated in D$p$-brane geometries with $p<5$.

Since the standard polygamy inequality of $\EW$ is violated, it is therefore natural to ask whether there exists a weaker version of this inequality, or a weaker version of the opposite (\ie\ monogamy) inequality. Here we can take some hints from quantum information theory where such weaker inequalities in mixed state entanglement measures have indeed been suggested. For instance, though entanglement of formation can not be freely shared between subsystems, however, there exists an upper bound on the sum of their entanglement of formation \cite{PhysRevA.89.034303}, \ie\ implying a weaker monogamy condition for the entanglement of formation. A similar weaker inequality exists for the entanglement of purification \cite{PhysRevA.91.042323}. Motivated by these conditions in quantum information theory, we propose the following weaker monogamy inequality for the EWCS
\be
  \EW(A:BC) + \frac{I(A:BC)}{2} \geq \EW(A:B) + \EW(A:C) \ .\label{NewEWineqality}
\ee
We check that this inequality is satisfied for all the holographic cases we study in this article. In particular, we verify (\ref{NewEWineqality}) both analytically and numerically for AdS$_3$, whereas numerical evidence is provided for its validity in higher dimensions and other backgrounds considered in this work. 

Although, even if a quantum entanglement measure is not monogamous, the same quantity raised to a power may be. For instance, measures such as the concurrence, entanglement for formation, and negativity do not obey monogamy, while their squared versions likely will \cite{PhysRevA.61.052306,PhysRevA.75.062308,PhysRevA.91.012339}. In this context, it becomes interesting to ask whether an analogous monogamy condition exists for the square of the EWCS, \ie\ 
\be
 (\EW(A:BC))^2 \geq (\EW(A:B))^2 + (\EW(A:C))^2 \ .  \label{EWsq}
\ee
We find that the answer to this question appears to be affirmative, at least for all the cases investigated in this article. The presence of this monogamy condition in turns leads us to a new lower bound for the $\EW(A:BC)$ which reads as follows 
\be\label{ewlb}
 \EW(A:BC)\geq \sqrt{(\EW(A:B))^2 + (\EW(A:C))^2} \ .
\ee
We also verify this lower bound for various bulk configurations. It is interesting to note that while the inequalities (\ref{EWsq}) and (\ref{ewlb}) are essentially mathematically equivalent, their physical interpretations differ.

In summary, our results  show that when we have a tripartite configuration, three disjoint slabs, then the EWCS is neither polygamous nor monogamous but rather weakly monogamous. The monogamy condition is obeyed by the square of the EWCS and this monogamy condition can be used to extract a new lower bound for $\EW(A:BC)$. The property of the EWCS relating it to the extensiveness of the mutual information and also to the strong subadditivity of the EE is preserved.

The rest of this article is organized as follows. In Sec.~\ref{sec:bounds} we will overview some basics on holographic EE and EWCS. Further, we will briefly talk about some existing holographic inequalities and in particular present novel inequalities for the EWCS. In the subsequent Sec.~\ref{sec:results} we test the inequalities in various setups. We end the article in Sec.~\ref{sec:discussion}
with a discussion and our conclusions.

\section{Bounding holographic entanglement measures}\label{sec:bounds}   

In this section we will first touch base with the computation of EE in holographic theories, based on the conjecture by RT~\cite{Ryu:2006bv}. We will then focus on the EWCS, discuss its definition, and overview the existing inequalities suggested to be obeyed by it if a connection to known information quantities is assumed. We point out the violation of an existing inequality and continue with proposing novel inequalities for the EWCS. In the subsequent section we test these inequalities in various cases.

\subsection{Holographic entanglement entropy}

EE is a measure of pure state entanglement and is expressed as the von Neumann entropy of the reduced density matrix $\rho_A$ of the subsystem $A$ 
\begin{equation}
 S(A) = -\mathrm{Tr}_A\rho_A\log\,\rho_A\ . \label{vne}
\end{equation}
To calculate the EE for field theories, one can use the replica method \cite{Calabrese:2009qy} but it has limitations. Holography provides another way to calculate the EE for field theories having known duals. The EE of a quantum field theory region $A$ bounded by $\partial A$ is proportional to the area of a minimal surface $\Gamma_{8}$ anchored to $\partial A$ on the spacetime boundary \cite{Ryu:2006bv,Ryu:2006ef}. $\Gamma_8$ is an eight-dimensional surface embedded in the ten-dimensional spacetime and thus the holographic EE is 
\be\label{eq:hee}
 S(A) = \frac{1}{4G_{10}}\int_{\Gamma_8} d^{8}x e^{-2\phi}\sqrt{\det g_s} \ .
\ee
In this expression the induced metric $g_s$ on $\Gamma_8$ with coordinates $x_\mu$, $\mu=0,\ldots,7$, is written in the string frame, $\phi$ is the dilaton, and $G_{10}$ is the ten-dimensional Newton constant. Notice that $\Gamma_8$ is the minimal surface, so it minimizes the expression on the right hand side. We only consider static configurations in this work, generalization of (\ref{eq:hee}) to time-dependent states has been discussed in \cite{Hubeny:2007xt}. For cases with the known field theory dual, the proposal (\ref{eq:hee}) is expected to give the leading $N_c^2$ order result in the large coupling limit. Extensions beyond leading order in either parameter are currently not well understood.

In many of the cases that we will consider, with the exception of those in Sec.~\ref{sec:Dpbackgrounds}, the dilaton is a constant. It is useful to pass from the string frame to the Einstein frame, which is obtained by sending the metric from $g_s$ to $g = e^{-\phi/2} g_s$.  
Notice that $\Gamma_8$ explores also the internal geometry, although in the cases that we will discuss in this work this proceeds trivially. The bulk surface $\Gamma_8=\Gamma_A\times {\cal M}$ is factorized in $(d-1)$-dimensional part $\Gamma_A$ in the external dimensions and a compact part filling the whole $(10-d-1)$-dimensional internal manifold ${\cal M}$. The end result of the integral in the internal directions thus boils down to a multiplicative constant corresponding to the volume of the internal space, \ie\   \be
\frac{1}{G_N^{(d+1)}}= \frac{\int_{\cal M} d^{10-d-1}x \sqrt{\det g_{10-d-1}}}{G_{10}}\equiv  \frac{{\rm{Vol}{\cal M}}}{G_{10}} \ , 
\ee
where $g_{10-d-1}$ is the reduction of the induced metric on the internal space, $G_N^{(d+1)}$ is the Newton constant in $(d+1)$ bulk dimensions. Armed with this we will take as our operational definition, in the Einstein frame, for the holographic EE the following expression 
\begin{equation}
 S(A) = \frac{\mathcal{A}_{\Gamma_A}}{4G_N^{(d+1)}}\ \ , \ \   \mathcal{A}_{\Gamma_A} = \int_{\Gamma_A} d^{d-1}x\sqrt{\det g_{d-1}} \ , \label{eq:hee2}
\end{equation}
where $g_{d-1}$ is the induced metric on $\Gamma_A$, which is assumed to be minimal and homologous to the spacetime boundary $A$.

The holographic EE is known to obey certain inequalities like the subadditivity, strong subadditivity, Araki-Lieb, monogamy of mutual information, and others \cite{Ryu:2006ef,Headrick:2013zda,Hayden:2011ag,Headrick:2007km,Bao:2015bfa}. The RT conjecture has thus been established pretty firmly and so one could take this a step further, \ie\ one can contemplate that any holographic spacetime should obey them.
Further, the test of these inequalities can also shed light on the nature/sharing of correlations for a given holographic setup.

\subsection{Entanglement wedge cross section}\label{sec:EW}

Having established that the holographic proposal for the computation of an EE for pure states passes several non-trivial tests, we now continue to the main focus of this article which is the entanglement wedge.

EE is considered a good measure of pure state entanglement but falls short if the state is mixed. To this end, the EWCS was proposed as a better geometric candidate  \cite{Takayanagi:2017knl}. The dual interpretation of the EWCS is currently an open issue and hence in this article we focus mainly on the inequalities pertaining to the EWCS. We believe that our work might shed light on its dual interpretation, too.

In order to compute the EWCS we follow the method outlined in \cite{Takayanagi:2017knl}. We consider two non-intersecting subsystems $A$ and $B$ in $d$-dimensional boundary field theory. The minimal surfaces in the $d$-dimensional constant time slice in the bulk pertaining to $A$, $B$, and $AB=A\cup B$ are $\Gamma_A$, $\Gamma_B$, and $\Gamma_{AB}$, respectively. The $d$-dimensional  entanglement wedge $M_{AB}$ in the bulk is then expressed as a region bounded by $A$, $B$, and $\Gamma_{AB}$, \ie\ whose boundary satisfies the following 
\begin{equation}
 \partial M_{AB}=A\cup B\cup \Gamma_{AB} \ . \label{EWboundary}
\end{equation}
Notice that the entanglement wedge does not exist when $A$ and $B$ are widely separated relative to their sizes. 

\begin{figure}[h]
\centering
\begin{tikzpicture}
\begin{scope}[scale=1.6]
\draw [red, thick] (-13,0) -- (-11.5,0);
\draw (-11.5,0) -- (-10.5,0);
\draw [red, thick](-10.5,0) -- (-9,0);

\draw[red] (-13,0) .. controls (-12.75,-2.5) and (-9.25,-2.5) .. (-9,0);
\fill[blue!10!] (-13,0) .. controls (-12.75,-2.5) and (-9.25,-2.5) .. (-9,0);
\fill[white!100!] (-11.5,0) .. controls (-11.25,-0.75) and (-10.75,-0.75) .. (-10.5,0);
\draw[red] (-11.5,0) .. controls (-11.25,-0.75) and (-10.75,-0.75) .. (-10.5,0);
\draw[dashed] (-11,-0.6) -- (-11,-1.86);
\node()at (-11,-1.25){$\Sigma_{AB}$};
\node()at (-9.7,0.25){$A$};
\node()at (-9.7,-0.25){$l_2$};
\node()at (-9.8,-1.55){$\Gamma^{(A)}_{AB}$};
\node()at (-10.6,-0.3){$\Gamma^{(A)}_{AB}$};
\node()at (-12.2,-1.55){$\Gamma^{(B)}_{AB}$};
\node()at (-11.4,-0.3){$\Gamma^{(B)}_{AB}$};
\node()at (-11,0.15){$s_1$};
\node()at (-12.3,0.25){$B$};
\node()at (-12.3,-0.25){$l_1$};
\end{scope}
\end{tikzpicture}
\caption{Pictorial representation for $\EW(A:B)$. The region in the blue color is the entanglement wedge $M_{AB}$. The boundary of the wedge (the red curve) is obtained as the union of $A$, $B$, $\Gamma_{AB}^{(A)}$, and $\Gamma_{AB}^{(B)}$. The dashed curve represents $\Sigma_{AB}$.} 
\label{MAB}
\end{figure}
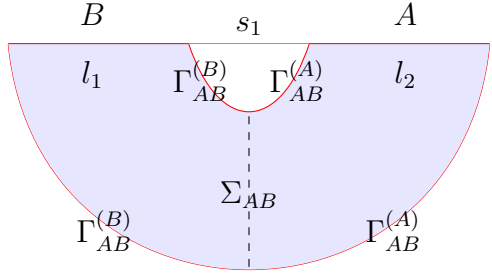

We now move on to divide $\Gamma_{AB}$ in two separate components in the following fashion 
\begin{equation}
 \Gamma_{AB} = \Gamma_{AB}^{(A)}\cup \Gamma_{AB}^{(B)} \label{RTdiv}
\end{equation}
such that 
\begin{eqnarray}
& & \tilde{\Gamma}_{A} = A\cup \Gamma_{AB}^{(A)}\,  \nonumber \\
& & \tilde{\Gamma}_{B} = B\cup \Gamma_{AB}^{(B)}\ .
\label{RTA}
\end{eqnarray}
Using the above Eqs. \eqref{RTdiv} and \eqref{RTA}, we can now divide the boundary of the entanglement wedge in the following way 
\begin{equation}
 \partial M_{AB}=\tilde{\Gamma}_{A}\cup \tilde{\Gamma}_{B} \ .
\end{equation}
One can now define the minimal surface $\Sigma_{AB}$, see Fig.~\ref{MAB},  which satisfies the following:\footnote{Recall that the second, homology condition means that $\Sigma_{AB}$ and $\tilde \Gamma_A$ together form the boundary of some $d$ dimensional domain inside $M_{AB}$. Notice that this condition is not symmetric under the exchange of $A$ and $B$, or at least this cannot be immediately verified. 
For the configurations considered in this article, however, the homology condition plays no role and the EWCS defined this way is indeed symmetric.}
\begin{equation}
\begin{split}
&(i)\ \partial\Sigma_{AB}=\partial\tilde{\Gamma}_{A}=\partial\tilde{\Gamma}_{B}\ ,\\
&(ii)\ \Sigma_{AB}\ \mathrm{is\ homologous\ to}
\ \tilde{\Gamma}_{A}\ \mathrm{inside}\ M_{AB} \ .
\end{split}
\end{equation}

We can now use the area of $\Sigma_{AB}$ to define the EWCS $E_{W}(A:B)$ as 
\begin{equation}\label{EW}
 \EW(A:B) = \min_{\Gamma_{AB}^{(A)}\subset\Gamma_{AB}}\left[\frac{\mathcal{A}(\Sigma_{AB})}{4G_N^{(d+1)}}\right] \ .
\end{equation}
To put in words, $\EW(A:B)$ is given by the minimal surface area of the cross section of the entanglement wedge $M_{AB}$ which connects the two subsystems $A$ and $B$.

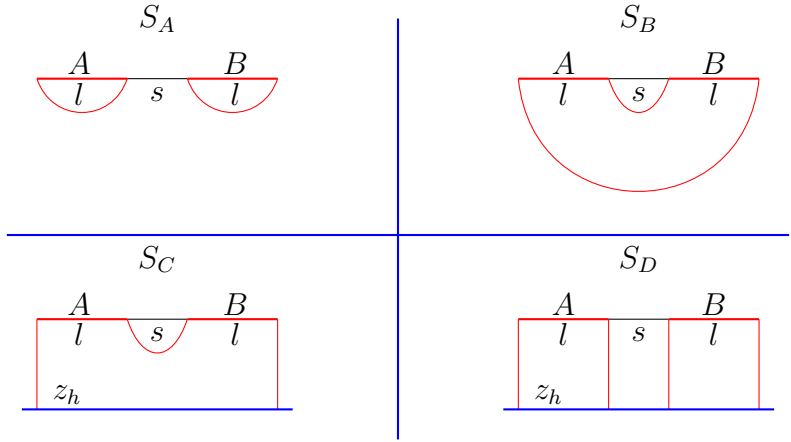
\begin{figure}[h]
\begin{center}
\begin{tikzpicture}
\begin{scope}[scale=0.8]
\draw [red, thick] (-13,0) -- (-11.5,0);
\draw (-11.5,0) -- (-10.5,0);
\draw [red, thick](-10.5,0) -- (-9,0);
\draw[red] (-13,0) .. controls (-12.75,-0.75) and (-11.75,-0.75) .. (-11.5,0);
\draw[red] (-10.5,0) .. controls (-10.25,-0.75) and (-9.25,-0.75) .. (-9,0);
\node()at (-9.7,0.25){$B$};
\node()at (-9.7,-0.25){$l$};
\node()at (-11,-0.25){$s$};
\node()at (-11,1){$S_A$};
\node()at (-12.3,0.25){$A$};
\node()at (-12.3,-0.25){$l$};
\draw [blue, thick](-13.5,-2.6) -- (-0.5,-2.6);
\draw [blue, thick](-7,1) -- (-7,-6);
\draw [red, thick] (-5,0) -- (-3.5,0);
\draw (-3.5,0) -- (-2.5,0);
\draw [red, thick](-2.5,0) -- (-1,0);
\draw[red] (-5,0) .. controls (-4.75,-2.5) and (-1.25,-2.5) .. (-1,0);
\draw[red] (-3.5,0) .. controls (-3.25,-0.75) and (-2.75,-0.75) .. (-2.5,0);
\node()at (-3,1){$S_B$};
\node()at (-4.25,0.25){$A$};
\node()at (-4.25,-0.25){$l$};
\node()at (-3,-0.25){$s$};
\node()at (-1.75,-0.25){$l$};
\node()at (-1.75,0.25){$B$};
\draw [red, thick] (-13,-4) -- (-11.5,-4);
\draw (-11.5,-4) -- (-10.5,-4);
\draw [red, thick](-10.5,-4) -- (-9,-4);
\draw[red] (-11.5,-4) .. controls (-11.25,-4.75) and (-10.75,-4.75) .. (-10.5,-4);
\draw[red] (-13,-4) -- (-13,-5.5);
\draw[red] (-9,-4) -- (-9,-5.5);
\draw [blue, thick](-13.25,-5.5) -- (-8.75,-5.5);
\node()at (-12.5,-5.25){$z_h$};
\node()at (-9.7,-3.75){$B$};
\node()at (-9.7,-4.25){$l$};
\node()at (-11,-4.25){$s$};
\node()at (-11,-3){$S_C$};
\node()at (-12.3,-3.75){$A$};
\node()at (-12.3,-4.25){$l$};
\draw [red, thick] (-5,-4) -- (-3.5,-4);
\draw (-3.5,-4) -- (-2.5,-4);
\draw [red, thick](-2.5,-4) -- (-1,-4);
\draw[red] (-5,-4) -- (-5,-5.5);
\draw[red] (-3.5,-4) -- (-3.5,-5.5);
\draw[red] (-1,-4) -- (-1,-5.5);
\draw[red] (-2.5,-4) -- (-2.5,-5.5);
\draw [blue, thick](-5.25,-5.5) -- (-0.75,-5.5);
\node()at (-4.5,-5.25){$z_h$};
\node()at (-3,-3){$S_D$};
\node()at (-4.25,-3.75){$A$};
\node()at (-4.25,-4.25){$l$};
\node()at (-3,-4.25){$s$};
\node()at (-1.75,-3.75){$B$};
\node()at (-1.75,-4.25){$l$};

\end{scope}
\end{tikzpicture}
\end{center}
\caption{Pictorial representation of the four different minimal surface configurations in the bulk for a bipartite configuration having two slabs of equal width $l$ separated by a distance $s$.} 
\label{ewbipartite}
\end{figure}


\begin{figure}[h]
\centering
    \includegraphics[width=0.7\textwidth]{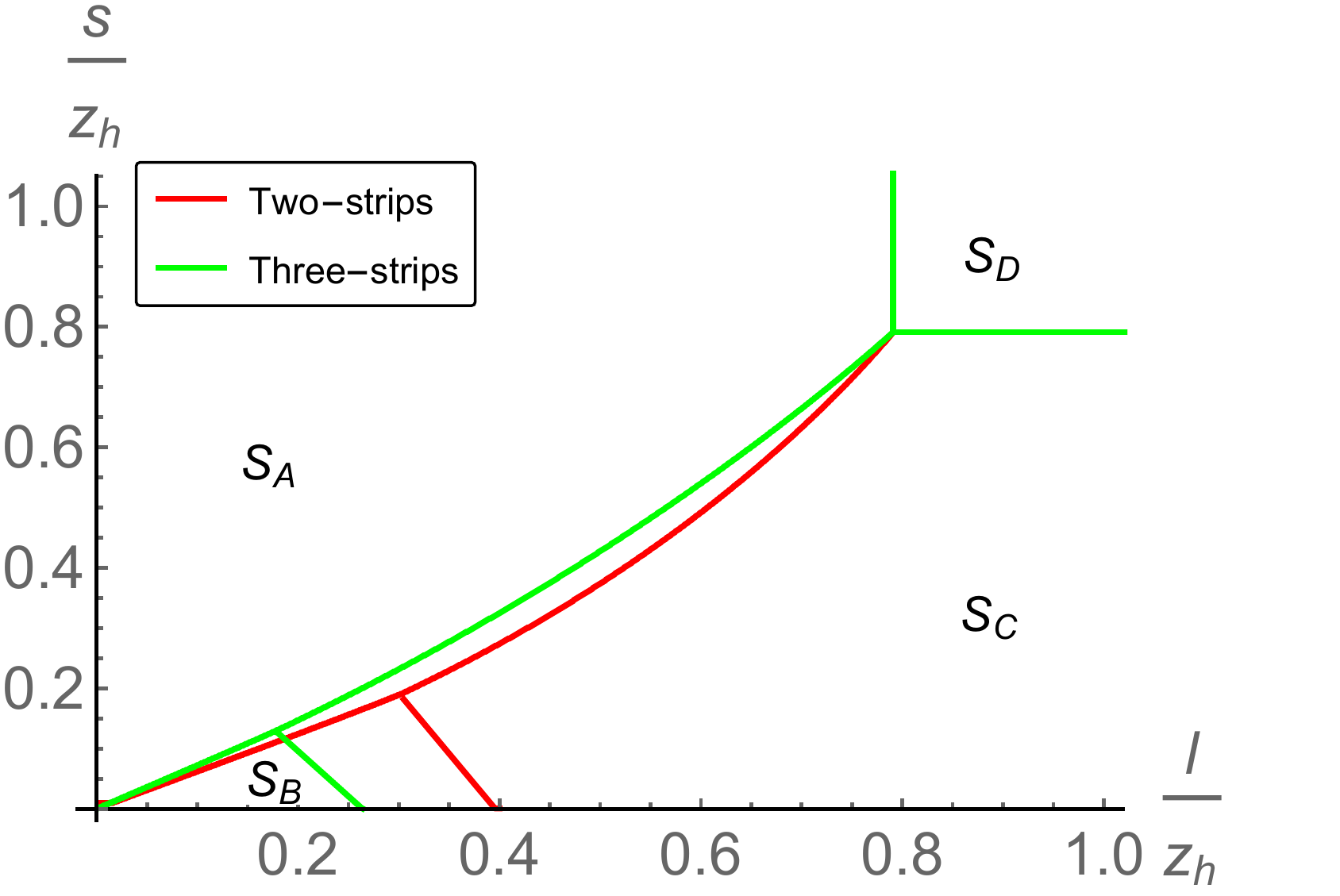}
    \caption{Phase diagram for the bipartite and tripartite configuration in the AdS soliton background in $d=3$.}
\label{fig:phase diag}
\end{figure}

\begin{figure}[h]
\centering
\begin{tikzpicture}
\draw [red, thick] (-5,0) -- (-3.5,0);
\draw (-3.5,0) -- (-2.5,0);
\draw [red, thick](-2.5,0) -- (-1,0);
\draw (-1,0) -- (0,0);
\draw [red, thick](0,0) -- (1.5,0);
\draw[red] (-5,0) .. controls (-4.75,-1) and (-3.75,-1) .. (-3.5,0);
\draw[red] (-2.5,0) .. controls (-2.25,-1) and (-1.25,-1) .. (-1,0);
\draw[red] (0,0) .. controls (0.25,-1) and (1.25,-1) .. (1.5,0);
\node()at (-4.25,0.25){$B$};
\node()at (-4.25,-0.25){$l_1$};
\node()at (-3,-0.25){$s_1$};
\node()at (-0.5,-0.25){$s_2$};
\node()at (-1.75,0.25){$A$};
\node()at (-1.75,-0.25){$l_2$};
\node()at (0.75,0.25){$C$};
\node()at (0.75,-0.25){$l_3$};
\end{tikzpicture}



\vspace{5mm}
\begin{tikzpicture}
\begin{scope}[xshift=-4cm]
\draw [red, thick] (-5,0) -- (-3.5,0);
\draw (-3.5,0) -- (-2.5,0);
\draw [red, thick](-2.5,0) -- (-1,0);
\draw (-1,0) -- (0,0);
\draw [red, thick](0,0) -- (1.5,0);
\draw[red] (-5,0) .. controls (-4.25,-2.5) and (0.75,-2.5) .. (1.5,0);
\draw[red] (-3.5,0) .. controls (-3.25,-0.75) and (-2.75,-0.75) .. (-2.5,0);
\draw[red] (-1,0) .. controls (-0.75,-0.75) and (-0.25,-0.75) .. (0,0);
\draw[dashed] (-2.9,-0.55) .. controls (-2.4,-1.1) and (-1.1,-1.1) .. (-0.6,-0.55);
\node()at (-1,-1.3){$\Sigma(A:BC)$};
\node()at (-4.25,0.25){$B$};
\node()at (-4.25,-0.25){$l_1$};
\node()at (-3,-0.25){$s_1$};
\node()at (-0.5,-0.25){$s_2$};
\node()at (-1.75,0.25){$A$};
\node()at (-1.75,-0.25){$l_2$};
\node()at (0.75,0.25){$C$};
\node()at (0.75,-0.25){$l_3$};
\end{scope}
\begin{scope}[xshift=4cm]
\draw [red, thick] (-5,0) -- (-3.5,0);
\draw (-3.5,0) -- (-2.5,0);
\draw [red, thick](-2.5,0) -- (-1,0);
\draw (-1,0) -- (0,0);
\draw [red, thick](0,0) -- (1.5,0);
\draw[red] (-5,0) .. controls (-4.25,-2.5) and (0.75,-2.5) .. (1.5,0);
\draw[red] (-3.5,0) .. controls (-3.25,-0.75) and (-2.75,-0.75) .. (-2.5,0);
\draw[red] (-1,0) .. controls (-0.75,-0.75) and (-0.25,-0.75) .. (0,0);
\draw[dashed] (-3.1,-0.55) .. controls (-3.2,-0.9) and (-3.4,-1.4) .. (-3.5,-1.47);
\draw[dashed] (-0.4,-0.55) .. controls (-0.3,-0.9) and (0,-1.4) .. (0.1,-1.47);
\node()at (-0.2,-0.95){$\Sigma(A:BC)$};
\node()at (-3.3,-0.95){$\Sigma(A:BC)$};
\node()at (-4.25,0.25){$B$};
\node()at (-4.25,-0.25){$l_1$};
\node()at (-3,-0.25){$s_1$};
\node()at (-0.5,-0.25){$s_2$};
\node()at (-1.75,0.25){$A$};
\node()at (-1.75,-0.25){$l_2$};
\node()at (0.75,0.25){$C$};
\node()at (0.75,-0.25){$l_3$};
\end{scope}
\end{tikzpicture}



\begin{tikzpicture}
\draw [red, thick] (-13,0) -- (-11.5,0);
\draw (-11.5,0) -- (-10.5,0);
\draw [red, thick](-10.5,0) -- (-9,0);
\draw[red] (-11.5,0) .. controls (-11.25,-0.75) and (-10.75,-0.75) .. (-10.5,0);
\draw[red] (-13,0) .. controls (-12.75,-2.5) and (-9.25,-2.5) .. (-9,0);
\draw[dashed] (-11,-0.6) -- (-11,-1.86);
\node()at (-11,-2.25){$\Sigma(A:B)$};
\node()at (-9.7,0.25){$A$};
\node()at (-9.7,-0.25){$l_2$};
\node()at (-11,-0.25){$s_1$};
\node()at (-12.3,0.25){$B$};
\node()at (-12.3,-0.25){$l_1$}; 
\draw [red, thick] (-5,0) -- (-3.5,0);
\draw (-3.5,0) -- (-2.5,0);
\draw [red, thick](-2.5,0) -- (-1,0);
\draw[red] (-5,0) .. controls (-4.75,-2.5) and (-1.25,-2.5) .. (-1,0);
\draw[red] (-3.5,0) .. controls (-3.25,-0.75) and (-2.75,-0.75) .. (-2.5,0);
\draw[dashed] (-3,-0.6) -- (-3,-1.86);
\node()at (-3,-2.25){$\Sigma(A:C)$};
\node()at (-4.25,0.25){$A$};
\node()at (-4.25,-0.25){$l_2$};
\node()at (-3,-0.25){$s_2$};
\node()at (-1.75,0.25){$C$};
\node()at (-1.75,-0.25){$l_3$};
\end{tikzpicture}
\caption{Pictorial representations of the configurations in $S_A$ and $S_B$ phases for the case of three disjoint intervals with subsystem $A$ in the middle. \textbf{Top row:} the tripartite configuration in the $S_A$ phase. No EWCSs exist in this case. \textbf{Middle row:} the tripartite configuration in the $S_B$ phase. The EWCS is obtained through the area of the $\Sigma(A:BC)$ (dashed curves) either from the standard single component configuration (left) or the double component configuration (right). \textbf{Bottom row:} Bipartite configurations in the $S_B$ phase.  The dashed curves in the left and right plots represent $\Sigma(A:B)$ and $\Sigma(A:C)$, respectively.} 
\label{SAB}
\end{figure}
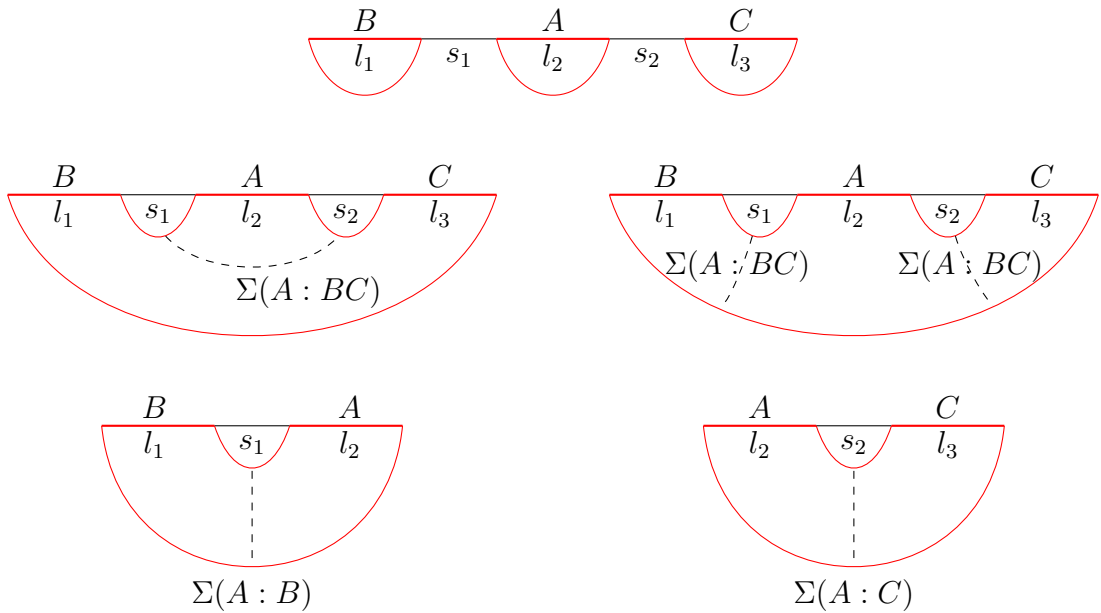

\begin{figure}[h]
\centering
\begin{tikzpicture}
\begin{scope}[xshift=-4.2cm]
\draw [red, thick] (-5,0) -- (-3.5,0);
\draw (-3.5,0) -- (-2.5,0);
\draw [red, thick](-2.5,0) -- (-1,0);
\draw (-1,0) -- (0,0);
\draw [red, thick](0,0) -- (1.5,0);
\draw [red, thick] (-5,0) -- (-5,-2);
\draw [red, thick] (1.5,0) -- (1.5,-2);
\draw[red] (-3.5,0) .. controls (-3.25,-0.75) and (-2.75,-0.75) .. (-2.5,0);
\draw[red] (-1,0) .. controls (-0.75,-0.75) and (-0.25,-0.75) .. (0,0);
\draw[dashed] (-2.9,-0.55) ..  controls (-2.4,-1.1) and (-1.1,-1.1) .. (-0.6,-0.55);
\draw [blue] (-5.5,-2) -- (2,-2);
\node()at (-4,-1.8){$z_h$};
\node()at (-1,-1.3){$\Sigma(A:BC)$};
\node()at (-4.25,0.25){$B$};
\node()at (-4.25,-0.25){$l_1$};
\node()at (-3,-0.25){$s_1$};
\node()at (-0.5,-0.25){$s_2$};
\node()at (-1.75,0.25){$A$};
\node()at (-1.75,-0.25){$l_2$};
\node()at (0.75,0.25){$C$};
\node()at (0.75,-0.25){$l_3$};
\end{scope}
\begin{scope}[xshift=4.2cm]
\draw [red, thick] (-5,0) -- (-3.5,0);
\draw (-3.5,0) -- (-2.5,0);
\draw [red, thick](-2.5,0) -- (-1,0);
\draw (-1,0) -- (0,0);
\draw [red, thick](0,0) -- (1.5,0);
\draw [red, thick] (-5,0) -- (-5,-2);
\draw [red, thick] (1.5,0) -- (1.5,-2);
\draw[red] (-3.5,0) .. controls (-3.25,-0.75) and (-2.75,-0.75) .. (-2.5,0);
\draw[red] (-1,0) .. controls (-0.75,-0.75) and (-0.25,-0.75) .. (0,0);
\draw[dashed] (-3,-0.6) -- (-3,-2);
\draw[dashed] (-0.5,-0.6) -- (-0.5,-2);
\draw [blue] (-5.5,-2) -- (2,-2);
\node()at (-4,-1.8){$z_h$};
\node()at (-2.85,-1.3){$\Sigma(A:BC)$};
\node()at (-0.35,-1.3){$\Sigma(A:BC)$};
\node()at (-4.25,0.25){$B$};
\node()at (-4.25,-0.25){$l_1$};
\node()at (-3,-0.25){$s_1$};
\node()at (-0.5,-0.25){$s_2$};
\node()at (-1.75,0.25){$A$};
\node()at (-1.75,-0.25){$l_2$};
\node()at (0.75,0.25){$C$};
\node()at (0.75,-0.25){$l_3$};
\end{scope}
\end{tikzpicture}
\vspace{8mm}


\begin{tikzpicture}
\draw [red, thick] (-13,0) -- (-11.5,0);
\draw (-11.5,0) -- (-10.5,0);
\draw [red, thick](-10.5,0) -- (-9,0);
\draw[red] (-11.5,0) .. controls (-11.25,-0.75) and (-10.75,-0.75) .. (-10.5,0);
\draw [red, thick] (-13,0) -- (-13,-2);
\draw [red, thick] (-9,0) -- (-9,-2);
\draw [blue] (-13.5,-2) -- (-8.5,-2);
\draw[dashed] (-11,-0.6) -- (-11,-2);
\node()at (-12,-1.8){$z_h$};
\node()at (-11,-2.4){$\Sigma(A:B)$};
\node()at (-9.7,0.25){$A$};
\node()at (-9.7,-0.25){$l_2$};
\node()at (-11,-0.25){$s_1$};
\node()at (-12.3,0.25){$B$};
\node()at (-12.3,-0.25){$l_1$};
\draw [red, thick] (-5,0) -- (-3.5,0);
\draw (-3.5,0) -- (-2.5,0);
\draw [red, thick](-2.5,0) -- (-1,0);
\draw [red, thick] (-5,0) -- (-5,-2);
\draw [red, thick] (-1,0) -- (-1,-2);
\draw[red] (-3.5,0) .. controls (-3.25,-0.75) and (-2.75,-0.75) .. (-2.5,0);
\draw[dashed] (-3,-0.6) -- (-3,-2);
\draw [blue] (-5.5,-2) -- (-0.5,-2);
\node()at (-4,-1.8){$z_h$};
\node()at (-3,-2.4){$\Sigma(A:C)$};
\node()at (-4.25,0.25){$A$};
\node()at (-4.25,-0.25){$l_2$};
\node()at (-3,-0.25){$s_2$};
\node()at (-1.75,0.25){$C$};
\node()at (-1.75,-0.25){$l_3$};
\end{tikzpicture}
\vspace{3mm}



\begin{tikzpicture}
\draw [red, thick] (-5,0) -- (-3.5,0);
\draw [red, thick](-2.5,0) -- (-1,0);
\draw [red, thick](0,0) -- (1.5,0);

\draw [red, thick] (-5,0) -- (-5,-2);
\draw [red, thick] (-3.5,0) -- (-3.5,-2);
\draw [red, thick] (-1,0) -- (-1,-2);
\draw [red, thick] (-2.5,0) -- (-2.5,-2);
\draw [red, thick] (0,0) -- (0,-2);
\draw [red, thick] (1.5,0) -- (1.5,-2);
\draw [blue] (-5.5,-2) -- (2,-2);
\draw (-3.5,0) -- (-2.5,0);
\draw (-1,0) -- (0,0);
\node()at (-4,-1.8){$z_h$};
\node()at (-4.25,0.25){$B$};
\node()at (-4.25,-0.25){$l_1$};
\node()at (-3,-0.25){$s_1$};
\node()at (-0.5,-0.25){$s_2$};
\node()at (-1.75,0.25){$A$};
\node()at (-1.75,-0.25){$l_2$};
\node()at (0.75,0.25){$C$};
\node()at (0.75,-0.25){$l_3$};
\end{tikzpicture}
\caption{Pictorial representations of the configurations in $S_C$ and $S_D$ phases, appearing in soliton geometries, for the case of three non-overlapping subsystems with $A$ in the middle. Note that none of the surfaces really end on different points (all blue points at $z_h$ are identified) but they are connected $\sqcup$-embeddings. \textbf{Top row:} the tripartite configuration in the $S_C$ phase. The EWCS is obtained through the area of the $\Sigma(A:BC)$ (dashed curves) either from the standard single component configuration (left) or the double component configuration (right). \textbf{Middle row:} Bipartite configurations in the $S_C$ phase.  The dashed curves in the left and right plots represent $\Sigma(A:B)$ and $\Sigma(A:C)$, respectively.
\textbf{Bottom row:} the tripartite configuration in the $S_D$ phase. No EWCSs exist in this phase. }
\label{SCD}
\end{figure}
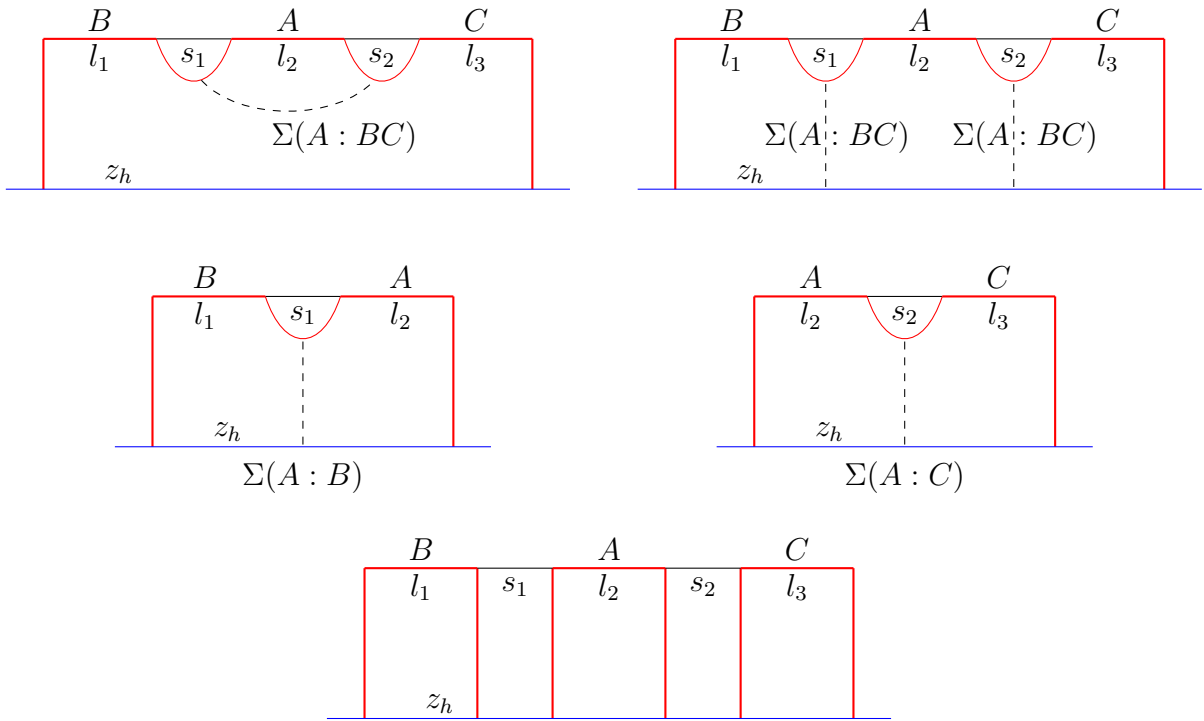

\subsection{Setup: tripartite and bipartite configurations}\label{sec:configs}

Having defined the quantities we will be interested in this work, we move to the discussion of the geometries and configurations that we will study. We will be studying various geometries (pure AdS, AdS black hole, AdS soliton, and D$p$-brane geometries) that can be interpreted to be duals of various field theories in Minkowski space. That is, we consider (pure) AdS in the Poincar\'e patch rather than global AdS, for example.

For our purpose, we will focus on the three disjoint interval configurations on the boundary, meaning that we have three subsystems $A,B,C$ separated by finite distances. After labeling, we can have two inequivalent choices, \ie\ we can have the subsystem $A$ either on the side or in the middle. We will primarily focus on the configuration where $A$ is placed in the middle as it leads to non-trivial results, assuming the ``standard'' labeling of the three sets in the inequalities. Notice, however, that even if the case of $A$ in the side will lead to simpler structure of the inequalities, it will have some important consequences for the polygamy and monogamy properties of the EWCSs, as we will discuss below.

One needs to be careful to pick the correct minimal configuration for the entanglement entropy. For bipartite systems, there are four possibilities which are shown in Fig.~\ref{ewbipartite}. 
Of these four ``phases'', $S_A$ and $S_B$ appear in all geometries that we will study, whereas the phases $S_C$ and $S_D$ appear only in the soliton geometries. The same phases can also be defined analogously for tripartite systems, as we shall show below. 
As an example, the phase diagram for bipartite and tripartite configurations for $d=3$ for the AdS$_4$ soliton background is shown in Fig.~\ref{fig:phase diag}. Here, we have considered the symmetric configurations, where all the subsystems have equal widths $l$ and are separated by an equal distance $s$. Though the phase diagram is drawn only for $d=3$, the presence of four phases $\{S_A, S_B, S_C, S_D\}$ is a generic feature for any $d\geq3$ in the AdS soliton background. 

Assuming the symmetric case where all strips have the same widths and the separations between them are also equal, the expressions of these four extremal surfaces are given by 
\begin{eqnarray}
& &  S_A = m S(l)\ ,~~~ S_B= (m-1) S(s) + S(ml+(m-1)s) \ , \nonumber \\
& & S_C = (m-1) S(s) + S_\sqcup\ ,~~~ S_D = m S_\sqcup \,.
\end{eqnarray}
Here $m=2(3)$ for the bipartite (tripartite) case and $S(l)$ and $S_\sqcup$ represent the entanglement entropy of short U-shaped and long ``rectangular" $\sqcup$-surfaces\footnote{This extremal surface is approximately three flat pieces glued together: two of them extend from the UV all the way down to the IR, whereas the third one lays at the bottom of the geometry connecting the two.} with one strip. Equating the expressions in the above equations will give the phase boundaries and elsewhere one needs to pick the one which gives minimal configuration, \ie\ depending upon the size and the separation between the subsystems these four phases exchange dominance and some may not even exist.  It is important to note that for pure AdS, AdS black brane, and non-conformal D$p$-brane background we will only have the $S_A$ and $S_B$ phases. This phase diagram dictates which configuration of the minimal surfaces is relevant for our purpose. In this article, we focus on the  
$S_B$ phase for pure AdS, AdS black brane, and non-conformal D$p$-brane geometries. For AdS soliton we consider the $S_B$ and $S_C$ phases. 
This is because in the other phases ($S_A$ of $S_D$) all the inequalities we shall study will become trivial. This is true even if we choose parameters such that the configurations for one of the relevant bipartite pairs, ($A$ and $B$) or ($A$ and $C$) enters one of these phases. 
For further discussion, see~\cite{Ben-Ami:2014gsa}.

The entanglement wedges do not exist in the $S_A$ phase. This is shown in the top row of Fig.~\ref{SAB}. We show here the configuration for the tripartite state, but since the minimal surfaces of the strips are independent, it is immediate that the EWCSs are absent for the bipartite pairs ($A$ and $B$) or ($A$ and $C$) as well. For the $S_B$ phase, the EWCS $\EW(A:BC)$ is shown in the middle row, and the wedges $\EW(A:B)$ and $\EW(A:C)$ are shown in the bottom row of Fig.~\ref{SAB}.
In case of the $S_C$ phase, $\EW(A:BC)$ will be as shown in the top row of Fig.~\ref{SCD} while $\EW(A:B)$ and $\EW(A:C)$ are shown in the middle row. The entanglement wedges do not exist in the $S_D$ phase as well and this phase is shown in the bottom row of Fig.~\ref{SCD}.
Notice that for tripartite configurations in the $S_B$ phase (middle row in Fig.~\ref{SAB}) and in the $S_C$ phase (top row in Fig.~\ref{SCD}), there are two possible configurations for the minimal surface $\Sigma(A:BC)$. As it follows from the definitions, the configuration with the lowest area determines the value of the EWCS in this case. In most of the setups that we study, the minimal area is given by the left ``standard'' configuration where the surface  $\Sigma(A:BC)$ has a single component. However, for some lengths and separations of the strips in the AdS black hole and AdS soliton geometries, the right configurations (with two separate pieces) will contribute.

\subsection{Existing holographic inequalities for the entanglement wedge}

In this subsection we will talk about certain existing inequalities in relation to the EWCS that are relevant to our purpose. The EWCS $\EW(A:B)$ is always nonnegative and is free from UV divergences when the distance between $A$ and $B$ is nonzero. As previously stated, when the two subsystems $A$ and $B$ are far apart then $\EW(A:B) = 0$. In this case also the mutual information  $I(A:B)=S(A)+S(B)-S(A:B)$ 
is zero. On the other hand when the EWCS exists, then we have 
$I(A:B)>0$. Th EWCS and the holographic mutual information satisfy the following inequality 
\begin{equation}
 \EW(A:B)\geq \frac{1}{2}I(A:B) \ . \label{EWMI} 
\end{equation}
For a geometric proof of the above inequality (\ref{EWMI}), see \cite{Takayanagi:2017knl}. In the case of a pure state, the inequality (\ref{EWMI}) is saturated.

When we have a tripartite configuration on the boundary, \ie\ three non-intersecting subsystems $A,B,C$, then we have~\cite{Takayanagi:2017knl}
\begin{equation}
 \EW(A:BC)\geq \EW(A:B) \label{EWtri} 
\end{equation}
which is analogous to 
the property of the quantum mutual information~\cite{Bagchi_2015} 
\begin{equation}\label{MItri} 
 I(A:BC)\geq I(A:B) \ . 
\end{equation}
Further, the inequality (\ref{polygamycondition}),  the polygamy of the EWCS which we reproduce here for ease of reference is 
\begin{equation}\label{ewpoly} 
 \EW(A:B)+\EW(A:C)\geq \EW(A:BC) \ . 
\end{equation}
In this article, one of our main objectives is to test these $\EW$ inequalities in various settings. In particular, we present evidence that the inequality (\ref{EWtri}) is preserved and the inequality (\ref{ewpoly}) is violated in pure AdS, AdS black brane, AdS soliton, and D$p$-brane backgrounds.

\subsection{Novel inequalities for the entanglement wedge}\label{sec:ineqs}
On top of discussing the existing inequalities, we also present novel inequalities for $\EW$. These inequalities are motivated from quantum information theory and might play an important role in enhancing our understanding of $\EW$ and its field theory interpretation. 

The proposed new inequalities for $\EW$ are as follows 
\begin{itemize}
\item[1.]  
The inequality \begin{equation}
\EW(A:BC) + {I(A:BC)}/{2}\geq \EW(A:B) + \EW(A:C) \ .
\end{equation}
This inequality primarily declares that the EWCS is neither polygamous nor monogamous. Rather it is weakly monogamous. Let us reiterate that a quantum information measure (in our case $\EW$) for a given tripartite state is called monogamous if it satisfies  
\be\label{ewmono}
 \EW(A:BC) \geq \EW(A:B) + \EW(A:C)
\ee
and if it does not then it is polygamous. In other words, the reversed inequality (\ref{ewpoly}) holds. This monogamy property basically implies that the entanglement cannot be freely shared between the subsystems. 

As we found that the polygamy of $\EW$ is not satisfied in general, it is interesting to ask whether a variant of the monogamy inequality can be satisfied for $\EW$. In order to construct such a weak monogamy inequality one needs to add a finite, divergence free quantity to $\EW(A:BC)$ so that together they will serve as an upper bound to $\EW(A:B) + \EW(A:C)$. Since we are dealing with a tripartite configuration, the simplest possible choice would be to add the tripartite mutual information \ie\ $I(A:B:C)=I(A:B) + I(A:C) - I(A:BC)$. Recalling that for RT surfaces in general there exists a robust inequality $I(A:B:C)\leq 0$ \cite{Hayden:2011ag}, prompts us to posit the following 
\begin{eqnarray} \label{TriMIIneq}
 \EW(A:BC) - \frac{I(A:B:C)}{2} \geq \EW(A:B) + \EW(A:C)\ .
\end{eqnarray}
We found that such an inequality is violated for small $s/l$ ratio for $d\geq4$ already in case of pure AdS. Taking hints from Eqs.~(\ref{EWMI}) and (\ref{MItri}) and the fact that $I(A:BC)$ is positive in the relevant 
$S_B$ and $S_C$ phases, the next possible and simplest choice would be
\begin{eqnarray}\label{MIIneq}
 \EW(A:BC) + \frac{I(A:BC)}{2} \geq \EW(A:B) + \EW(A:C) \ .
\end{eqnarray}
We find (as will be discussed below) that the above Eq.~\eqref{MIIneq} is satisfied in all the gravitational duals that we have studied in this article. Interestingly, we do find a violation in this inequality for certain configurations corresponding to the $S_B$ phase, but values of slab widths and separations for which the EE is in the $S_A$ phase (so that we are computing in the ``wrong'' phase). In analogy to usual thermodynamics nomenclature, we call such configurations the ``subdominant'' $S_B$ phase, and similarly we call the true $S_B$ phase the ``dominant'' $S_B$ phase. Here one should note also that the transition from $S_B$ phase to some other phase (in our analysis typically the $S_A$ phase) takes place at different parameter values depending on which state we are studying (\eg\ the tripartite state or one of the bipartite pairs, $A$ and $B$ or $A$ and $C$). The relevant transition which determines the end of the true ``$S_B$ phase for the inequality'' is whichever of these happens first. 

Notice also that the inequalities~\eqref{TriMIIneq} and~\eqref{MIIneq} differ only by the term $(I(A:B) + I(A:C) )/2$ on the left hand side. That is, as the bipartite mutual informations are positive by definition,~\eqref{MIIneq} is weaker than~\eqref{TriMIIneq}: if~\eqref{TriMIIneq} held,~\eqref{MIIneq} would follow. Further, combining~\eqref{MIIneq} and~\eqref{EWMI}, we obtain
\be\label{EWMIIneq}
2 \EW(A:BC) \geq \EW(A:BC) + \frac{I(A:BC)}{2} \geq \EW(A:B) + \EW(A:C) \ ,
\ee
where the inequality between the first and the last expression can also be obtained from~\eqref{EWtri}. That is, the inequality~\eqref{MIIneq} is stronger than what is obtained by combining~\eqref{EWtri} for two different choices of the sets on the right hand side.

\item[2A.] The inequality \begin{equation}
(\EW(A:BC))^2\geq (\EW(A:B))^2 + (\EW(A:C))^2 \ .
\end{equation}
This inequality is motivated from quantum information theory where in many systems a squared version of the entanglement measure can satisfy the monogamy property even if the entanglement measure itself does not. Example includes concurrence, entanglement for formation, and entanglement negativity \cite{PhysRevA.61.052306,PhysRevA.75.062308,PhysRevA.91.012339}. In this work we find that the above squared version of $\EW$ is also monogamous, at least for all the background geometries considered in this article.

\item[2B.]  The inequality \begin{equation}
\EW(A:BC)\geq \sqrt{(\EW(A:B))^2 + (\EW(A:C))^2} \ .
\end{equation}
This third inequality is simply another way to express the monogamy of the squared $\EW(A:BC)$ which in turn leads us to a new lower bound of the linear $\EW(A:BC)$.
\end{itemize}

We thoroughly investigate these properties in pure AdS, AdS black brane, AdS soliton, and for non-conformal D$p$-brane backgrounds and our results seem to be robust. Since the interpretation of $\EW$ from the quantum information perspective is currently an open issue, our results therefore might provide a useful route in the hunt for the correct field theory dual to $\EW$.

\section{Results}\label{sec:results}

In this section, we carry out a detailed analysis of the polygamy and monogamy properties of EWCSs for tripartite states in various geometries. That is, we check whether the EWCS is polygamous or monogamous for each configuration, and whether the inequalities proposed above in Sec.~\ref{sec:ineqs} are satisfied. These checks are mostly based on extensive numerical analysis, but in a few cases also analytic results can be used. We will give evidence for
\begin{enumerate}
 \item  The violation of the polygamy as given by  (\ref{polygamycondition}).
 \item  The preservation of the inequality (\ref{TakaIneq}).
 \item  The preservation of the weak monogamy as given by (\ref{NewEWineqality}).
 \item  The preservation of the monogamy for the squared version of $\EW(A:BC)$ as given by (\ref{EWsq}).
 \item  The preservation of the lower bound for the linear version of $\EW(A:BC)$ as given by (\ref{ewlb}).
\end{enumerate}
Specifically, we find examples of violation of the polygamy of EWCSs in all classes of geometries that we study, \ie\ in pure AdS, thermal AdS, AdS soliton, and D$p$-brane geometries. 

We start by considering pure AdS geometries, \ie\ we will assume that the external space is AdS$_{d+1}$ times an internal geometry which decouples from the calculation.\footnote{The known dual field theories include for $d=2+1$ the ${\cal N}=6$ Chern-Simons matter theory \cite{Aharony:2008ug} and for $d=3+1$ the ${\cal N}=1$ SU(N) super Yang-Mills theory \cite{Maldacena:1997re}. The corresponding internal spaces in ten dimensions are the complex projective  manifold ${\mathbb{CP}}^3$ and the five-dimensional Sasaki-Einstein manifold $X^5$, respectively.
}
The metric for pure AdS$_{d+1}$ in the Poincar\'e patch is 
\begin{equation}
 ds^2 = \frac{L_{\mathrm{AdS}}^2}{z^2} \left( -dt^2+ dz^2+ dx^2+ dx_{d-2}^2 \right) \ ,
\end{equation}
where the boundary is at $z=0$ and $L_{\mathrm{AdS}}$ is the AdS radius.

\subsection{Pure AdS$_{2+1}$ geometry}\label{sec:pure AdS3}

To set the stage, we begin with the simplest case of $d=2$, corresponding to pure AdS in $2+1$ dimensions. In this case, both analytic and numerical results are possible for $\EW$, thereby allowing us to make concrete observations about its inequality properties. 

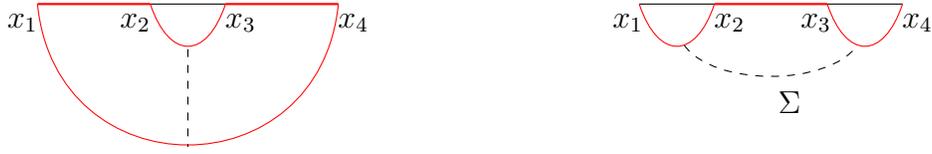
\begin{figure}[h]
\centering
\begin{tikzpicture}
\draw [red, thick] (-13,0) -- (-11.5,0);
\draw (-11.5,0) -- (-10.5,0);
\draw [red, thick](-10.5,0) -- (-9,0);
\draw[red] (-13,0) .. controls (-12.75,-2.5) and (-9.25,-2.5) .. (-9,0);
\draw[red] (-11.5,0) .. controls (-11.25,-0.75) and (-10.75,-0.75) .. (-10.5,0);
\draw[dashed] (-11,-0.6) -- (-11,-1.9);
\node()at (-9.7,0.25){\phantom{$l$}};
\node()at (-10.3,-0.25){$x_3$}; 
\node()at (-8.8,-0.25){$x_4$};
\node()at (-13.2,-0.25){$x_1$}; 
\node()at (-11.7,-0.25){$x_2$}; 

\draw (-5,0) -- (-4,0);
\draw [red, thick] (-4,0) -- (-2.5,0);
\draw (-2.5,0) -- (-1.5,0);
\draw[red] (-5,0) .. controls (-4.75,-0.75) and (-4.25,-0.75) .. (-4,0);
\draw[red] (-2.5,0) .. controls (-2.25,-0.75) and (-1.75,-0.75) .. (-1.5,0);
\draw[dashed] (-4.4,-0.55) .. controls (-4,-1.1) and (-2.5,-1.1) .. (-2.1,-0.55);
\node()at (-3,-1.3){$\Sigma$};
\node()at (-2.65,-0.25){$x_3$}; 
\node()at (-1.3,-0.25){$x_4$};
\node()at (-5.15,-0.25){$x_1$}; 
\node()at (-3.8,-0.25){$x_2$}; 
\end{tikzpicture}
\caption{Definitions for the analytic formulas for the EWCS in AdS$_3$. {\bf{Left:}} Before M\"{o}bius transformation. {\bf{Right:}} After M\"{o}bius transformation and renaming.} 

\label{MB}
\end{figure}

We consider the configuration where the subsystem $A$ is in the middle and, for simplicity, we consider the case where the subsystems $A$, $B$, and $C$ are of the same size and equally separated, \ie\ $l_1 = l_2 = l_3 = l$, and $s_1 = s_2 = s$. Later on, we will present results for the asymmetrical configuration where $s_1 \neq s_2$. The EWCS between two intervals has been computed for $d=2$ in \cite{Takayanagi:2017knl}, and is given by 
\begin{equation}
\EW 
= \frac{c}{6} \log\left( {1+2\xi+2\sqrt{\xi(\xi+1)}}\right) \ .
\label{EWAdS3}
\end{equation}
Here 
\begin{equation}
    c = \frac{3L_\mathrm{AdS}}{2G_N^{(3)}}
\end{equation}
is an overall coefficient (the central charge of the putative dual CFT) and $\xi$ is the cross ratio given by 
\begin{equation}\label{eq:crossratio}
\xi = \frac{(x_2-x_1)(x_4-x_3)}{(x_3-x_2)(x_4-x_1)} \ ,
\end{equation}
with the $x_i$ denoting the end points of the intervals in the parallel direction as shown in Fig.~\ref{MB} (left). To be precise, the EWCS is given by~\eqref{EWAdS3} when $\xi>1$ and vanishes when $\xi<1$. 
For our purpose, the configuration depicted in Fig.~\ref{SAB} (middle left), we need to compute a different minimal surface than that appearing in the EWCS of two intervals (shown in Fig.~\ref{MB} (right)). To do this we apply a trick:
we do a M\"{o}bius transformation that reorders the $x_i$'s cyclically such that $x_4$ is moved to the left of the other $x_i$'s, and rename the points such the original order is restored. As a result of this M\"{o}bius transformation, the EWSC configuration (that of Fig.~\ref{MB} (left)) is mapped to that shown in Fig.~\ref{MB} (right). Since the cross ratio is invariant under the M\"{o}bius transformation, and renaming the points takes $\xi\to{1}/{\xi}$, we find that $\EW \propto \mathrm{min}\,\mathcal{A}(\Sigma)$
for the types of surfaces shown in Fig.~\ref{MB} (right). For the (contribution to the) entanglement entropy, this gives
\begin{equation}
    \EW = \frac{c}{6} \log \left(\frac{\xi+2 +2 \sqrt{\xi +1}}{\xi }\right) \ . \label{anMobius} 
\end{equation}

For the configurations with equal geodesic lengths and separations, the cross ratios for various EWCS reduce to
\begin{equation}
\xi(A:B) = \xi(A:C) = \frac{1}{y(2+y)}
\end{equation}
where $y=s/l$. By substituting the above equations into Eq.~(\ref{EWAdS3}), we find the EWCSs $\EW(A:B)$ and $\EW(A:C)$. Similarly the relevant cross ratio for $\EW(A:BC)$ is
\begin{equation}
\xi(A:BC) = \frac{y^2}{2y+1} \ ,
\end{equation}
with the EWCS now obtained from~\eqref{anMobius} (or equivalently by substituting the inverse ratio in~\eqref{EWAdS3}).

Having found analytic expressions for all the EWCSs we next study their behavior with varying $y =s/l$ and also compare to numerical results.

\begin{figure}[h]
\begin{minipage}[c]{0.5\linewidth}
\includegraphics[width=1\textwidth]{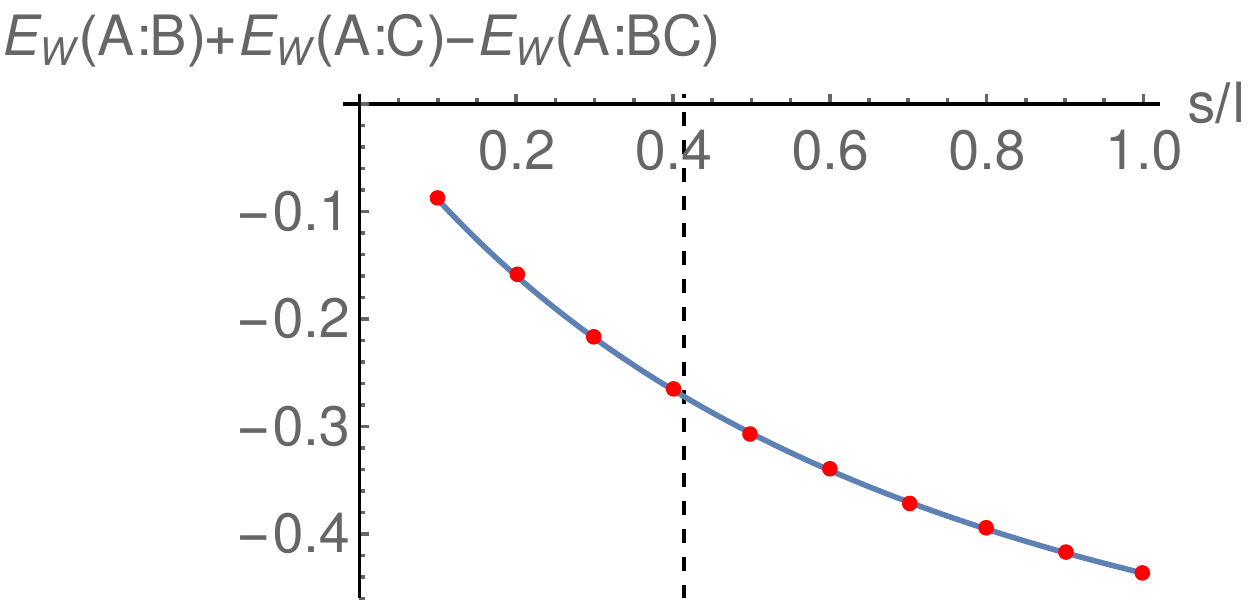}
\end{minipage}\hfill
\begin{minipage}[c]{0.5\linewidth}
\vspace{1cm}
\hspace{0.3cm}
\includegraphics[width=1\textwidth]{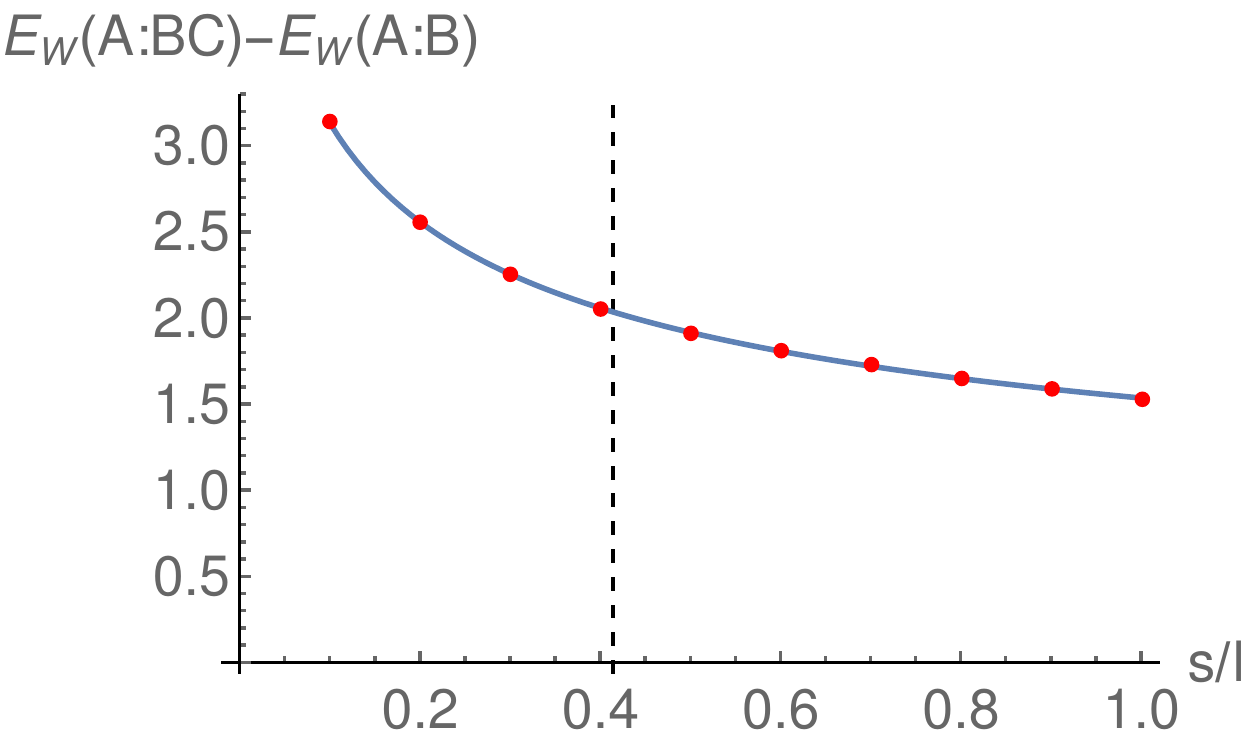}
\end{minipage}
\caption{{\bf{Left:}} $\EW(A:B) + \EW(A:C) - \EW(A:BC)$ for different $s/l$ ratios. {\bf{Right:}} $\EW(A:BC)- \EW(A:B)$ for different $s/l$ ratios. For both the plots, analytic result is depicted by a solid curve and the numerical result is shown by red dots. Bipartite critical point at $s/l \approx 0.414$ is indicated by a dashed curve.}
\label{fig:TakaIneqd2}
\end{figure}

\paragraph{The violation of the polygamy as given by (\ref{polygamycondition})}


The results for the quantity $\EW(A:B) + \EW(A:C) - \EW(A:BC)$ for different $s/l$ ratio is shown in Fig.~\ref{fig:TakaIneqd2} (left). In this figure, as well as in all the other plots in this subsection, we have set the proportionality coefficient $c/6 = L_\mathrm{AdS}/(4G_N^{(3)})$ to one. These results assume that we are in the $S_B$ phase, \ie\ $s/l$ is below the bipartite critical value, which is indicated by the vertical dashed black line. The bipartite critical point is the relevant transition point because its value is lower than the tripartite value (see Table~\ref{bicrit}), so it is the first value where any of the wedges in the inequality will change (and actually become zero).
If the plotted quantity is negative it would indicate a violation of the polygamy of $\EW$. From this figure, we  see that the value of this quantity can indeed be negative, both in the dominant (left side of the bipartite critical point) as well as in the subdominant (right side of the bipartite critical point) region of the $S_B$ phase, implying a clear violation of the claimed polygamy of $\EW$. We have further cross checked the same quantity numerically and the results are presented in Fig.~\ref{fig:TakaIneqd2}. The numerical results here are obtained using the general recipe presented in the subsection~\ref{sec:EW} and Appendices~\ref{app:surfaces} and~\ref{app:nummethod}. We see that our numerical results for $\EW(A:B) + \EW(A:C) - \EW(A:BC)$ correlate well with the analytical results, explicitly confirming both that $\EW$ does not always satisfy the polygamy inequality and that the numerical code works with high precision. We also remark that tripartite configurations where the strip $A$ is on the side rather than in the middle are monogamous as defined by~\eqref{monogamycondition}. We will discuss this in more detail in the case of AdS$_{d+1}$ geometries with generic $d$ below.  

\paragraph{The preservation of the inequality (\ref{TakaIneq})} 

Using the analytic results presented above, 
we can further analyze the inequality $\EW(A:BC) \geq \EW(A:B)$ in AdS$_{2+1}$. The behavior of this inequality for various values of the $s/l$ ratio is shown in Fig.~\ref{fig:TakaIneqd2} (right). We see that the value of $\EW(A:BC) - \EW(A:B)$ is always positive in the $S_B$ phase, implying that the inequality holds for $d = 2$. These results are further cross checked numerically and similar results are found.  

\begin{figure}[h]
\center
\includegraphics[scale=0.7]{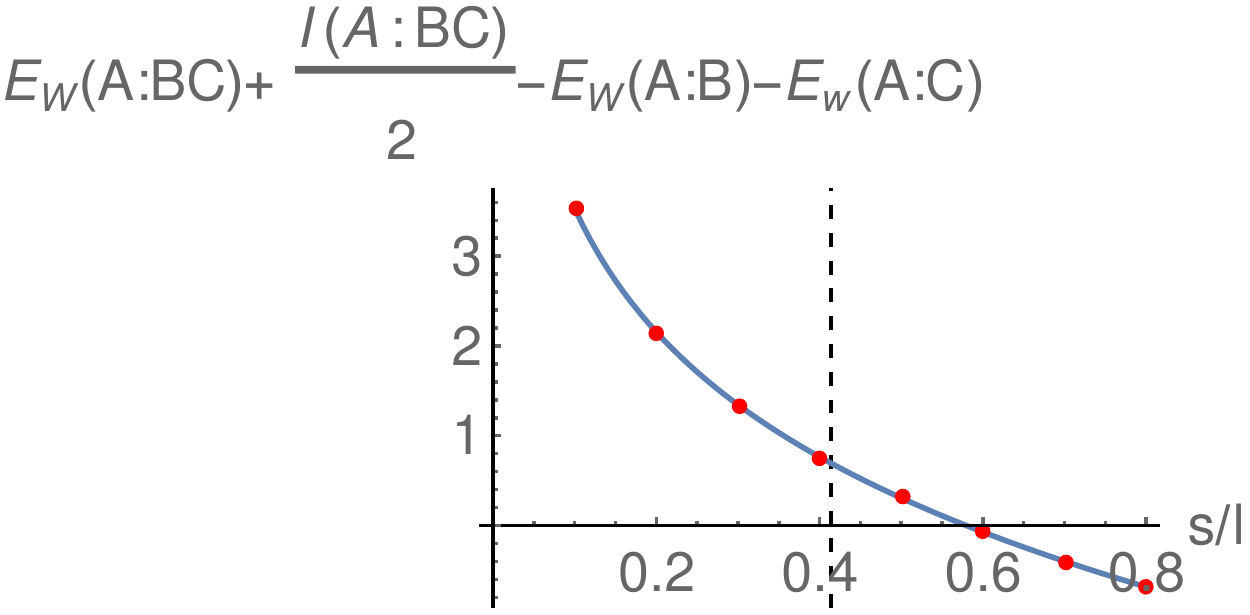}
\caption{$\EW(A:BC) + {I(A:BC)}/{2}- \EW(A:B) - \EW(A:C)$  for different $s/l$ ratios. Analytic result is depicted by a solid curve. Numerical result is shown by red dots. Bipartite critical point at $s/l \approx 0.414$ is indicated by a dashed line.}
\label{MIIneqd2} 
\end{figure}

\paragraph{The weak monogamy as given by (\ref{NewEWineqality})}
We now move on to discuss the weak monogamy inequality of $\EW$ described by  $\EW(A:BC) + I(A:BC)/2\geq \EW(A:B) + \EW(A:C)$. Here, $I(A:BC)$ is the holographic mutual information given by 
\begin{equation}
I(A:BC) = S(A) + S(BC)- S(ABC) \ .
\end{equation}
It can be further simplified to $I(A:BC)= 3S(l) - S(3l + 2s) - 2S(s)$ if $B$ and $C$ are far apart. Our analytic and numerical results for this inequality are presented in Fig.~\ref{MIIneqd2}. We observe that $\EW(A:BC) + I(A:BC)/2 - \EW(A:B) - \EW(A:C)$ is positive everywhere on the left side of the bipartite critical point. This indicates that the weak monogamy inequality is satisfied everywhere in the parameter space of $l$ and $s$ in the dominant $S_B$ phase. On the other hand, for larger $s/l$ values, this quantity can be negative, implying a violation of this weak monogamy. However, the large $s/l$ values correspond to the region where the $S_B$ phase is subdominant, and this violation therefore does not actually appear in the physical phase of the system.

\begin{figure}[h]
\centering
\includegraphics[width=0.49\textwidth]{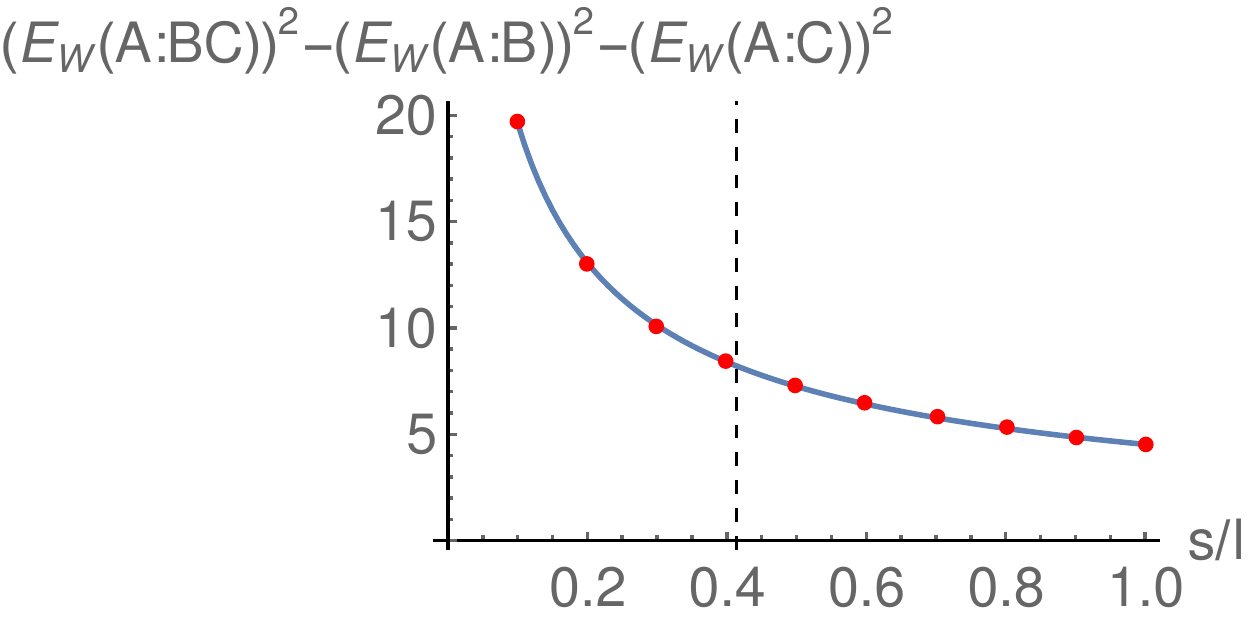}
\includegraphics[width=0.49\textwidth]{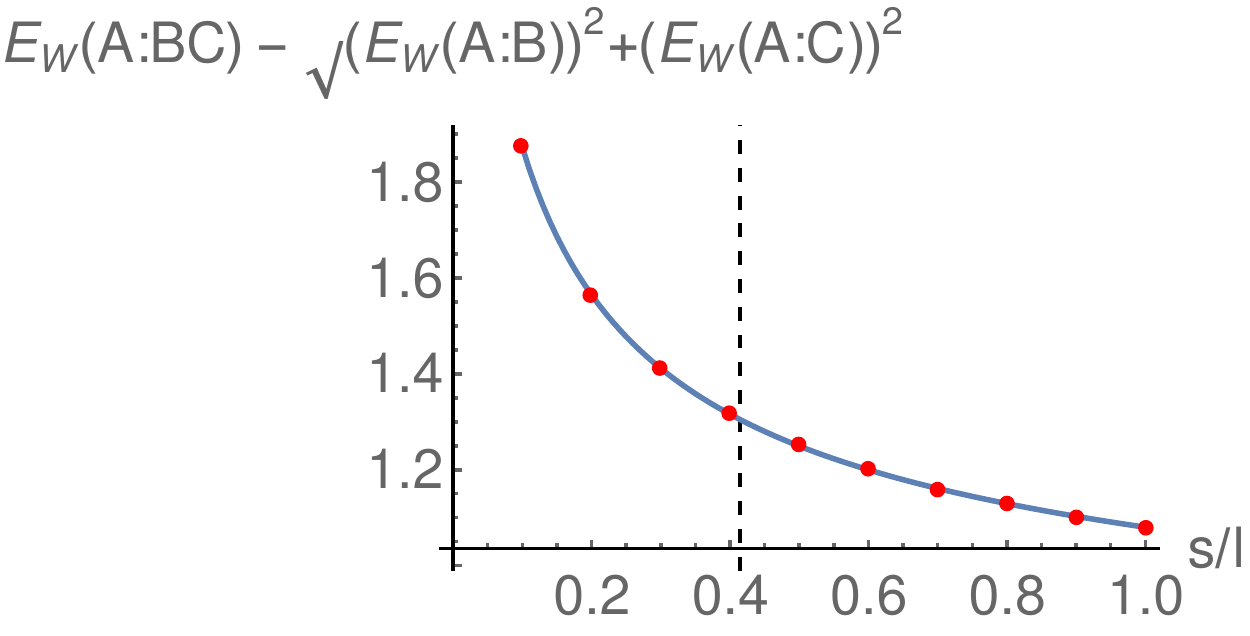}
\caption{{\bf{Left:}} $(\EW(A:BC))^2 - (\EW(A:B))^2 - (\EW(A:C))^2$ for different $s/l$ ratios. {\bf{Right:}} $\EW(A:BC)-\sqrt{(\EW(A:B))^2+(\EW(A:C))^2}$ for different $s/l$ ratios. For both the plots, analytic result is depicted by a solid curve and the numerical result is shown by red dots. Bipartite critical points at $s/l \approx 0.414$ are indicated by dashed lines in the panes.} \label{fig:SqIneqd2}
\end{figure}

\paragraph{The monogamy for the squared version of EWCS as given by (\ref{EWsq})}
We now show that the squared version of $\EW$, given by the inequality $(\EW(A:BC))^2\geq (\EW(A:B))^2 + (\EW(A:C))^2$, is monogamous. Our analytic and numerical results are shown in Fig.~\ref{fig:SqIneqd2}. We see that the quantity $(\EW(A:BC))^2 - (\EW(A:B))^2 - (\EW(A:C))^2$ is positive everywhere in the parameter space of $s$ and $l$. In particular, 
this quantity is positive in both the dominant as well as the subdominant region of the $S_B$ phase. Our numerical results again fit perfectly well with the analytic results, thereby demonstrating that the squared version of $\EW$ is indeed monogamous.

\paragraph{The lower bound for EWCS as given by (\ref{ewlb})}
The lower bound for the linear version of $\EW(A:BC)$ can be obtained from the monogamy property of $(\EW(A:BC))^2$ and is given by
(\ref{ewlb}). Our results for the lower bound are shown in Fig.~\ref{fig:SqIneqd2}. In agreement with the results for the squared inequality, we find that $\EW(A:BC)$ is always greater than $\sqrt{(\EW(A:B))^2 + (\EW(A:C))^2}$, suggesting that the latter invariably place constraints on $\EW(A:BC)$ and puts a lower bound on it.

\subsection{Pure AdS$_{d+1}$ geometry for generic $d$}\label{sec:pure AdS}
Having thoroughly discussed the various inequalities involving $\EW$ in AdS$_{2+1}$, we now move on to discuss them for generic $d$. Unfortunately, for $d>2$ analytic results for $\EW$ are hard to obtain, and we therefore have to rely on numerics to investigate these inequalities for a generic $d$. The numerical algorithm for $\EW$ is entirely analogous to the $d=2$ case, and is discussed in Appendix~\ref{app:nummethod}.

\begin{figure}[h]
\centering
\includegraphics[width=0.8\textwidth]{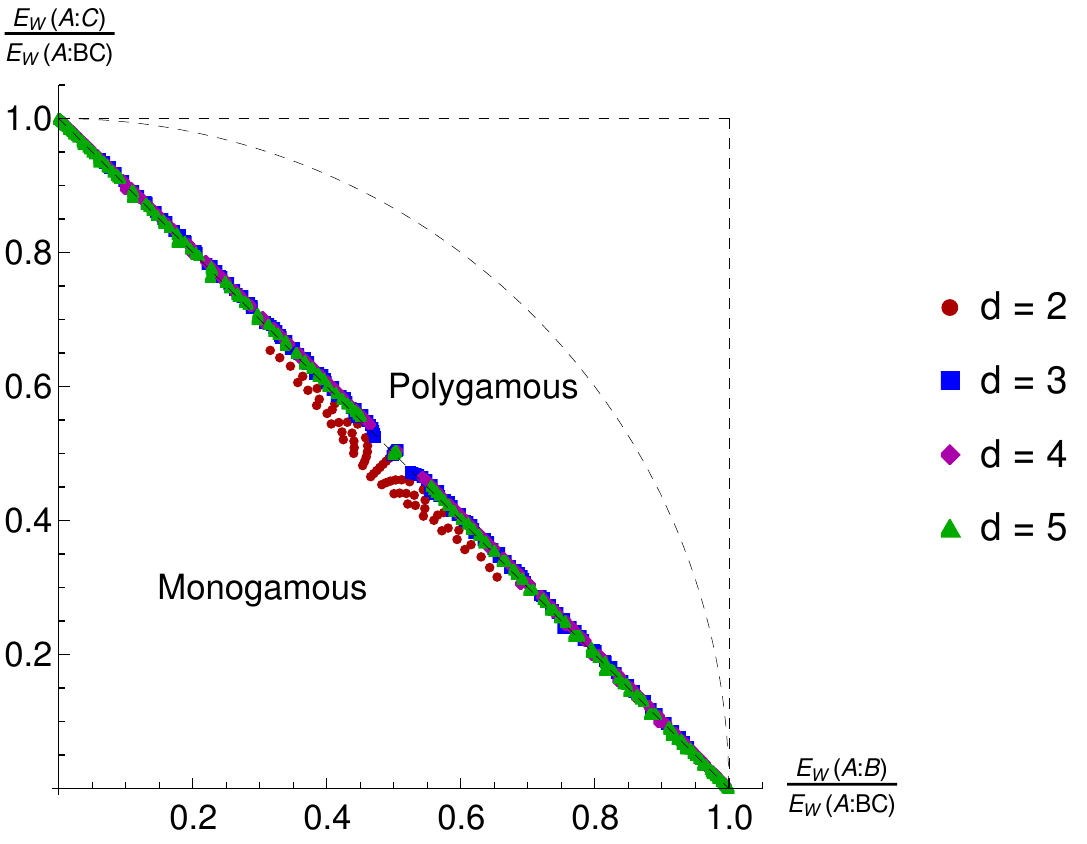}
\caption{
The ratio of the wedges $\EW(A:B)/\EW(A:BC)$ and  $\EW(A:C)/\EW(A:BC)$ in the $S_B$ phase for the pure AdS$_{d+1}$ geometries for various separations $s_1$ and $s_2$ of the intervals.
    }
\label{fig:asymmetricads}
\end{figure}

We begin by studying the EWCSs for the tripartite configuration, with $A$ in the middle, and assuming that all strip lengths are the same but letting the separations $s_1$ and $s_2$ vary independently.  
The ratios of the ECWSs from such configurations in the $S_B$ phase are shown in Fig.~\ref{fig:asymmetricads}.
That is, the data points were obtained by scanning over the ratios $s_1/l$ and $s_2/l$ over a grid that extends to the whole $S_B$ phase (meaning the true phase, not subdominant configurations).
In this plot, the dashed circular arc indicates the squared saturated equation $(\EW(A:BC))^2 = (\EW(A:B))^2 + (\EW(A:C))^2$ whereas the dashed inclining line indicates the equation $\EW(A:B) + \EW(A:C) = \EW(A:BC)$. 

Let us first comment what these results imply for the polygamy inequality (\ref{polygamycondition}).
As the inclined dashed line marks the saturation of this inequality, it allows to separate the region into polygamous and monogamous parts. Notice that for $d=2$, there are points which are below the inclined line, hence indicating monogamous behavior (\ref{ewmono}). These are the same results mentioned earlier for $d=2$, however, now in the asymmetrical configuration as well. For higher dimensions, the data points start clustering about the line $\EW(A:B) + \EW(A:C) = \EW(A:BC)$, and move towards the polygamy region. 

\begin{figure}[h]
\centering
    \includegraphics[width=0.8\textwidth]{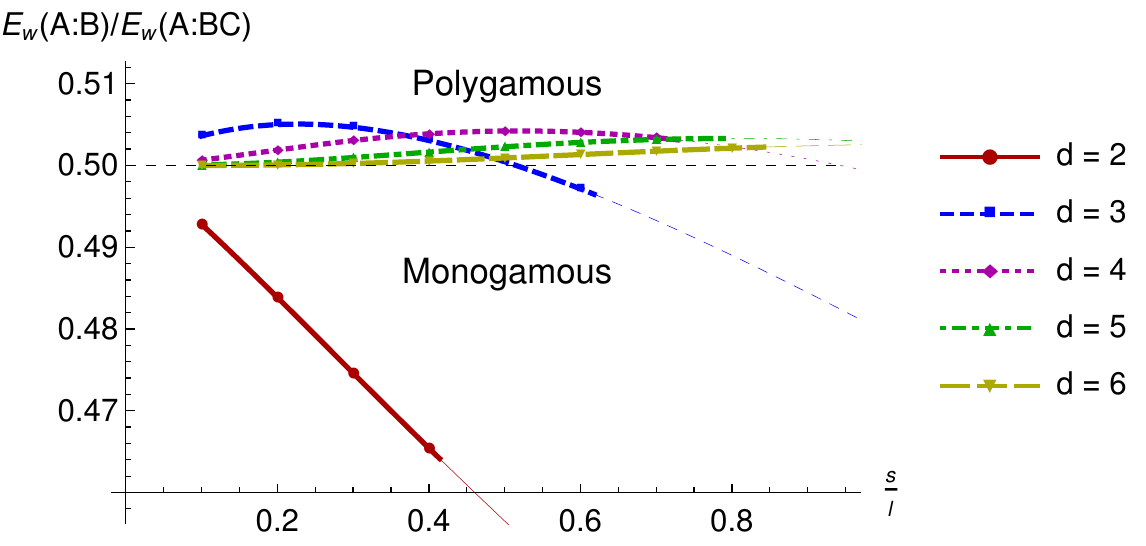}
    \caption{
The ratio $\EW(A:B)/\EW(A:BC)$ in the $S_B$ phase for the pure AdS$_{d+1}$ geometries as a function of $s/l$ in the symmetric configuration. The thick curves and dots show results in the region where the $S_B$ phase is dominant whereas the thin curves show the results in the region where the $S_B$ phase is subdominant.
    }
    \label{fig:symmetricads}
\end{figure}

In order to see the behavior more clearly, we plot the ratios for symmetric configurations, $s=s_1=s_2$, in Fig.~\ref{fig:symmetricads}, including the dependence on $s/l$. Due to symmetry, the two ratios of Fig.~\ref{fig:asymmetricads} coincide. In this plot, in addition to the true, dominant $S_B$ phase, we show the ratios in the subdominant region as thin curves. The transition from thick to thin curves is determined by the bipartite critical values given in Table~\ref{bicrit}.
We find a violation of the polygamy inequality in the dominant $S_B$ phase in $d=3$, whereas a similar violation is found in the subdominant $S_B$ phase in $d=4$. Notice also that the curve for $d=3$ crosses the horizontal line where the ratios equal $0.5$, indicating that the ECWS is neither polygamous nor monogamous for $d=3$. 

Notice that Figs.~\ref{fig:asymmetricads} and~\ref{fig:symmetricads} only contain the data for the tripartite configuration where $A$ is in the middle, and not the configuration where $A$ is on the side. If $A$ is on the side (and $B$ is in the middle), $A$ and $C$ are always so far apart that their entanglement wedges are separate and $\EW(A:C)=0$. In this case the known inequality~\eqref{TakaIneq} (which we will also explicitly verify below) implies that the monogamy condition~\eqref{monogamycondition} is satisfied. Moreover, since the entanglement wedges are nontrivial, saturating this inequality is not possible in general, so that the polygamy condition~\eqref{polygamycondition} is indeed violated. This holds independently of the number of dimensions.

Our results for the polygamy and monogamy properties for pure AdS spaces in different dimensions are therefore summarized as follows:
\begin{itemize}
\item The EWCS is monogamous in $d=2$, even including subdominant $S_B$ configurations.
\item The behavior of the EWCS changes from polygamous to monogamous as $s/l$ increases in $d=3$ (in the true dominant $S_B$ phase with tripartite configurations having $A$ in the middle). 
\item The EWCS shows polygamous behavior in the $S_B$ true, dominant phase in $d=4$ for tripartite states with $A$ in the middle. However, it is monogamous for part of subdominant $S_B$ phase and for configurations with $A$ on the side.  
\item The EWCS shows polygamous behavior in the $S_B$ true, dominant phase in $d \geq 5$ for tripartite states with $A$ in the middle. However, it is monogamous for configurations with $A$ on the side.  
\end{itemize}
Therefore, the only clean, configuration independent statement that we can make about the polygamy or monogamy properties is that the EWCS is monogamous for $d=2$. For $d \geq 3$ the EWCS is not polygamous nor monogamous in general.

The preservation of the inequalities  (\ref{TakaIneq}) and \eqref{EWsq} can also be read off from Figs.~\ref{fig:asymmetricads} and~\ref{fig:symmetricads}.
As all the data points are within the unit box in Fig.~\ref{fig:asymmetricads}, the ratios are always 
less than one, indicating the preservation of the inequality $\EW(A:BC) \geq \EW(A:B)$. In particular, it is satisfied in the dominant as well as in the subdominant $S_B$ configurations (as seen from Fig.~\ref{fig:symmetricads}) and remains so in all spacetime dimensions. 

Also notice from Fig.~\ref{fig:symmetricads} that, irrespective of the values of $s/l$ and $d$, the ratio $\EW(A:B)/\EW(A:BC)$ is always less than $1/\sqrt{2}$. Similarly, the data points in Fig.~\ref{fig:asymmetricads} are within the unit circle, denoted by the dashed curve. This result means that the squared version of inequality (\ref{EWsq}) is invariably satisfied everywhere in the parameter space both for the symmetric and asymmetric configurations, regardless of whether the $S_B$ phase is dominant there or not. Moreover, these results can be further used to put a lower bound on $\EW(A:BC)$ via the proposed inequality (\ref{ewlb}). In particular, for the symmetric configuration we always have $\EW(A:BC)\geq \sqrt{2}\EW(A:B)$. Recall that this lower bound is slightly higher than the lower bound coming from the inequality (\ref{TakaIneq}). Our results therefore suggest that the linear inequality relation and the lower bound of $\EW(A:BC)$ are stronger than previously anticipated in the literature. 

Next, we investigate the weak monogamy inequality of $\EW$ given by (\ref{NewEWineqality}). For this purpose, we define the following two quantities
\be\label{eq:RB}
 R_B = \frac{\EW(A:B)-\frac{1}{4}I(A:BC)}{\EW(A:BC)} \ , \qquad  R_C = \frac{\EW(A:C)-\frac{1}{4}I(A:BC)}{\EW(A:BC)}
\ee
in terms of which the weak monogamy inequality maps to $R_B+R_C \leq 1$. Notice that for the symmetric configuration $s_1=s_2$ we have $R_B=R_C$ so that the inequality  simplifies to $R_B\leq 1/2$. 

\begin{figure}[h]
    \centering
    \includegraphics[width=0.48\textwidth,trim= 0in -0.5in 0in 0in]{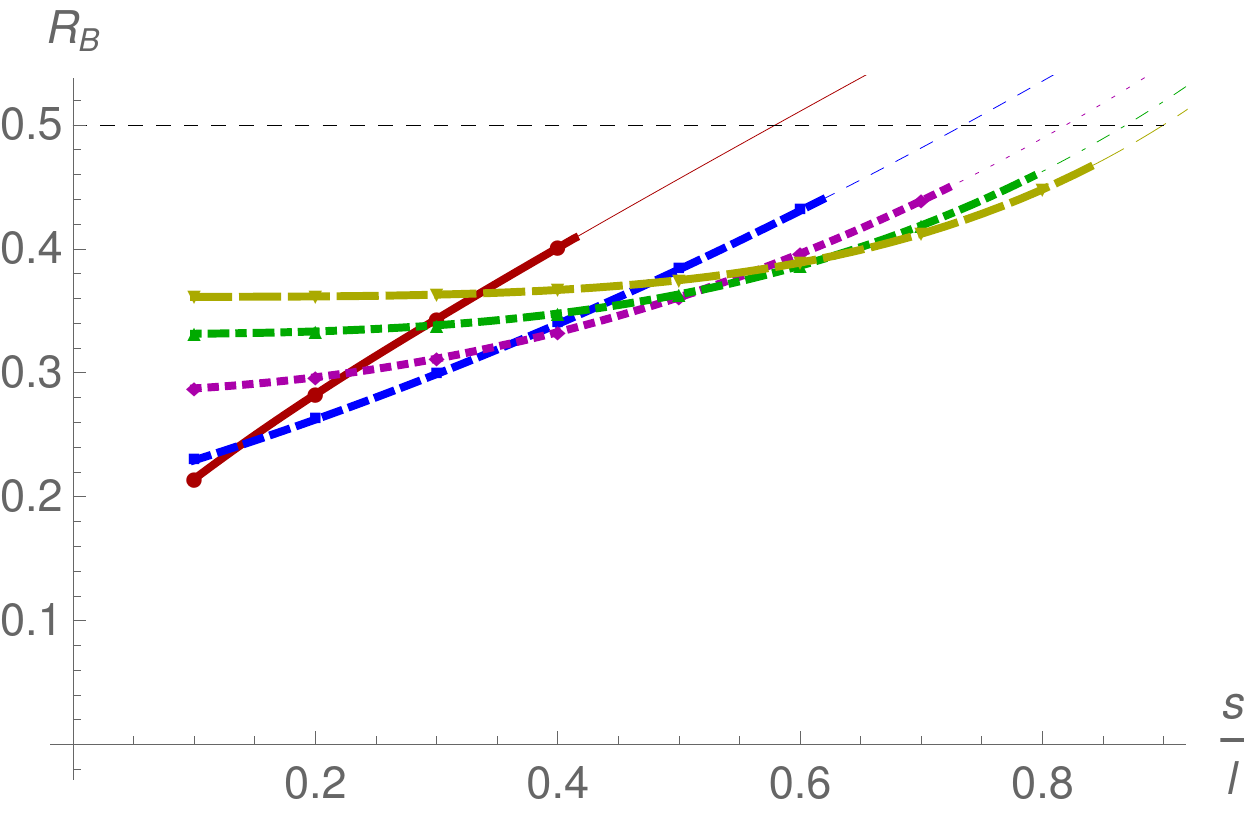}%
    \hspace{5mm}\includegraphics[width=0.46\textwidth]{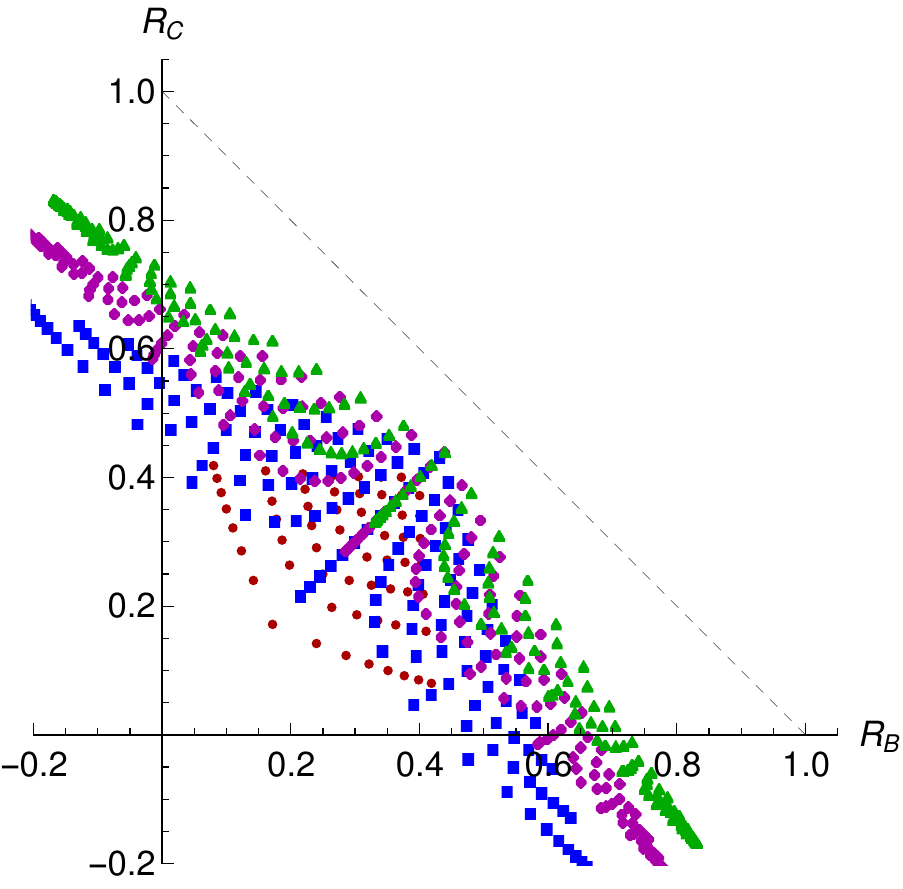}
    \caption{Demonstration of weak monogamy for pure AdS$_{d+1}$ geometries. {\bf{Left:}} The ratio for the symmetric configuration as a function of $s/l$. {\bf{Right:}} The ratio for the asymmetric configuration with different $s_1$ and $s_2$. Color scheme used here is similar to Fig.~\ref{fig:symmetricads}.
    }\label{fig:MI_ineq}
\end{figure}

Our results for the weak monogamy inequality are shown in Fig.~\ref{fig:MI_ineq}. In the left pane, the results are presented for the symmetric configuration whereas in the right pane the results are presented for the asymmetric configuration. These figures indicate that this inequality is preserved up to the bipartite critical point for the respective $d$ in the dominant $S_B$ phase. Above the critical point, \ie\ in the subdominant phase, we find violation in all dimensions.

\begin{table}
\begin{center}
\begin{tabular}{ |c|c|c| } 
 \hline
  & Bipartite critical point & Tripartite critical point \\
 \hline
$d$ = 2 & $\sqrt{2}-1 \approx 0.414$ & 1/2 \\ 
 \hline
 $d$ = 3 & $\frac{\sqrt{5}-1}{2} \approx 0.618$ & $\frac{\sqrt{10}-1}{3} \approx 0.721$ \\ 
 \hline
 $d$ = 4 & $\sqrt{3}-1 \approx 0.732$ & $\frac{\sqrt{7}-1}{2} \approx 0.823$ \\ 
 \hline
 $d$ = 5 & 0.800 & 0.874 \\ 
 \hline
 $d$ = 6 & 0.843 & 0.904 \\ 
 \hline
\end{tabular}
\vspace{0.2cm}\\
\caption{\label{bicrit}Bipartite and tripartite critical points for a symmetric configuration for various $d$ in case of pure AdS. Note the amusing connection to metallic numbers \cite{Balasubramanian:2018qqx}. See Appendix~\ref{app:critvals} for the derivation of the values.}
\end{center}
\end{table}

\subsection{AdS$_{d+1}$ soliton}

In this subsection, we present the results of $\EW$ inequalities in the AdS soliton geometry. The AdS soliton metric in $d+1$ bulk dimensions is
\be
 ds^2 = \frac{L_{\mathrm{AdS}}^2}{z^2} \left( -dt^2+ \frac{dz^2}{b(z)} + dx^2 + d\vec{x}_{d-3}^2 + b(z) dx_\text{circle}^2\right) \ ,
\ee
wherein the last spatial coordinate is compactified on a circle, $b(z)=1-\frac{z^d}{z_h^d}$, and the end point of the geometry $z_h$ is connected to the radius of the circle. 

For simplicity, we mainly concentrate on the symmetric configuration $l_1 = l_2 = l_3 = l$ and $s_1 = s_2 = s$ here, and the analogous results for the asymmetric configuration can be straightforwardly obtained. The phase diagram for bipartite and tripartite configuration for $d=3$ is shown in Fig.~\ref{fig:phase diag}. The bipartite and tripartite phase diagrams remain qualitatively the same in other dimensions, albeit with different critical lines.  

\begin{figure}[h]
\center
\includegraphics[width=0.8\textwidth]{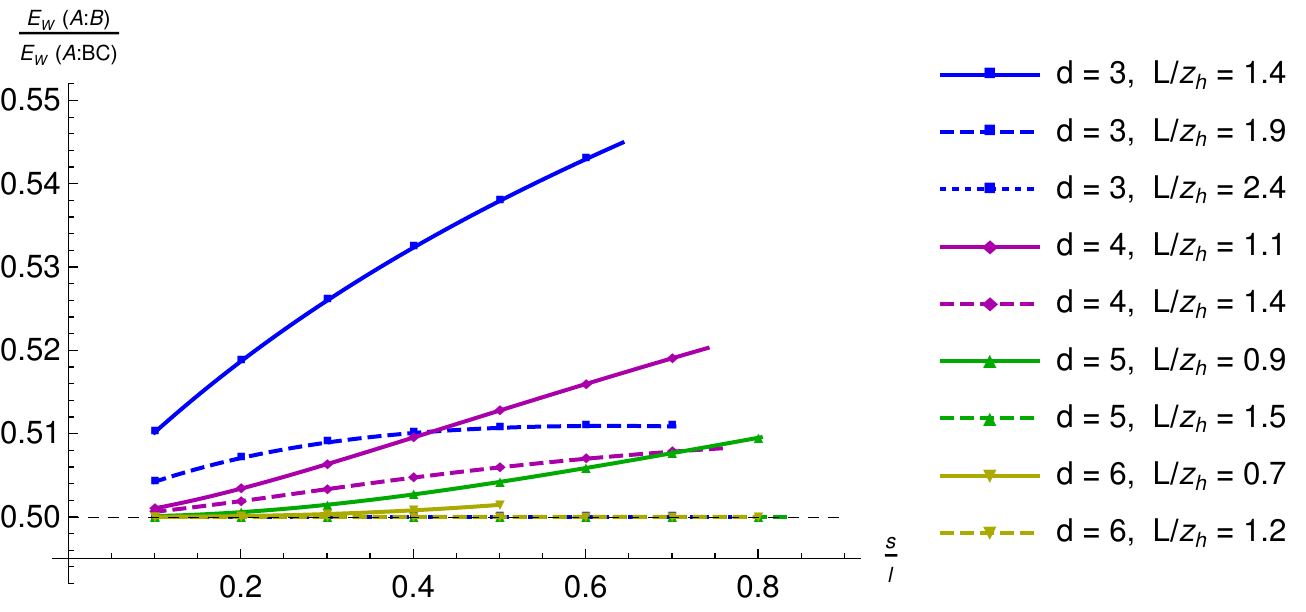}
\caption{
    The ratio of the wedges $\EW(A:B)/\EW(A:BC)$ in the $S_C$ phase in the AdS soliton background as a function of $s/l$ for various $d$ and total widths $L=3l+2s$.
    }
\label{fig:SC_phase}
\end{figure}

In the AdS soliton background, a non-trivial wedge can exist both in the $S_B$ and in the $S_C$ phases. The results of $\EW$ in the $S_B$ phase are practically identical to the pure AdS case. This is mainly because the $S_B$ phase is only found for small lengths and the tip region of the soliton geometry is not probed by the corresponding bulk surfaces. Therefore, here we primarily focus on the $S_C$ phase. For the $S_C$ phase, the wedge $\EW(A:BC)$ is illustrated in the top row of Fig.~\ref{SCD} while the wedges $\EW(A:B)$ and $\EW(A:C)$ are illustrated in the middle row of Fig.~\ref{SCD}.

From Figs.~\ref{SAB} and \ref{SCD}, it is clear that the results for $\EW(A:BC)$ in the $S_C$ phase are quite similar to the $S_B$ phase, and therefore depends only on the connected minimal surfaces. On the other hand, the cross sections $\EW(A:B)$ and $\EW(A:C)$ require the rectangular $\sqcup$ minimal surface. It is useful to note at this point that $\EW(A:B)$ is higher in the $S_C$ phase compared to the $S_B$ phase. This can be recognized from the fact that $\EW(A:B)$ in the $S_B$ phase is  
\be
 \EW(A:B) \propto \frac{L_{\mathrm{AdS}}^{d-1}}{4 G_{N}^{(d+1)}} \int_{z_*(s)}^{z_*(2l+s)} \frac{dz}{z^{d-1}} \ ,
\ee
where $z_*(l)$ is the turning point of the RT surface for a strip of length $l$.  In the $S_C$ phase we find
\be
 \EW(A:B)\propto \frac{L_{\mathrm{AdS}}^{d-1}}{4 G_{N}^{(d+1)}} \int_{z_*(s)}^{z_h} \frac{dz}{z^{d-1}} \ .
\ee
Since the turning point $z_*(2l+s)$ of the $S_B$ phase surface is always less than the end point $z_h$ in the AdS soliton background, 
it implies that $\EW(A:B)$ is higher in the $S_C$ phase compared to the $S_B$ phase. This has the following important implications.  In particular, 
the right side of the polygamy inequality (\ref{polygamycondition}) remains the same in both $S_B$ and $S_C$ phases whereas the left hand side is relatively higher in the $S_C$ phase compared to the $S_B$ phase, thereby suggesting that there might be higher chances for the polygamy inequality for $\EW$ to hold in the $S_C$ phase compared to the $S_B$ phase. 

Our numerical results confirm this expectation. We find that $\EW$ is always polygamous in the $S_C$ phase. This is shown in Fig.~\ref{fig:SC_phase}, where the behavior of ratio $\EW(A:B)/\EW(A:BC)$ in various dimensions is illustrated. From this figure, we notice that the ratio $\EW(A:B)/\EW(A:BC)$ is always greater than half, as is required to satisfy the polygamy in the symmetric configuration. We further find that this ratio is also always less than $1/\sqrt{2}$. This result further makes sure that the inequality of Eq.~\eqref{TakaIneq} and the squared monogamy inequality of Eq.~\eqref{EWsq} are simultaneously satisfied in the $S_C$ phase. Moreover, it also shows that our proposal for the lower bound in~\eqref{ewlb} holds in the $S_C$ phase as well.

Interestingly, from Fig.~\ref{fig:SC_phase} we see that apart from the ratio of the EWCSs being larger than $1/2$, it is actually exactly $1/2$ for a number of configurations. This is explained by the behavior of minimal surfaces in Fig.~\ref{SCD}: if $\EW(A:BC)$ is given by the surface in the top right plot, it is immediate that $ \EW(A:B) + \EW(A:C) = \EW(A:BC)$ so that polygamy and monogamy conditions are saturated. Since the surface of the top right figure provides, in more general, a lower bound for $\EW(A:BC)$. Moreover, one can check that for the same values of lengths, for the configuration where $A$ is on the side, $\EW(A:C)=0$ and  $\EW(A:B) = \EW(A:BC)$ by similar arguments, so the conditions are similarly saturated. In this sense, we find that EWCS is polygamous in the $S_C$ phase. Interestingly, this is the only extended, well defined region among the geometries that we have studied where we observe polygamy. Notice, however, that in the $S_B$ phase the results for the soliton geometry are similar to those of pure AdS, and there is no polygamy nor monogamy, in general.
 
\begin{figure}
\centering
    \includegraphics[width=0.6\textwidth]{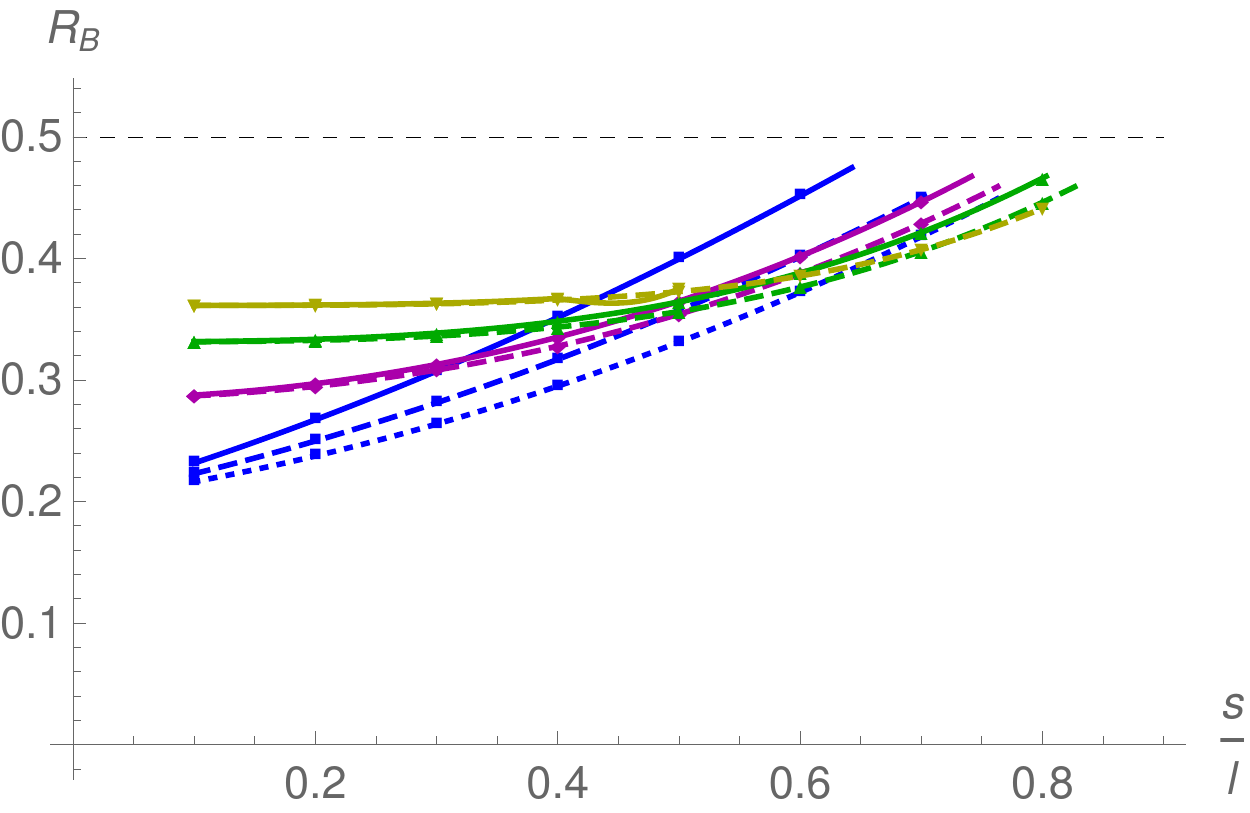}
    \caption{
    Demonstration of weak monogamy in the $S_C$ phase in the $(d+1)$-dimensional AdS soliton background as a function of $s/l$ for various $d$ and total widths $L=3l+2s$. Notation of curves as in Fig.~\ref{fig:SC_phase}.
    }\label{fig:SC_ineq}
\end{figure}

Lastly, the validity of the weak monogamy inequality~\eqref{NewEWineqality} up to the bipartite critical point is shown in Fig.~\ref{fig:SC_ineq}. The bipartite critical point here is the one that gives the transition from $S_C$ to the $S_A$ phase or to the $S_D$ phase. From the results in Fig.~\ref{fig:SC_ineq} we learn that $R_B$ (\ref{eq:RB}) is always less than $1/2$, confirming the validity of this inequality in the $S_C$ phase.  

\subsection{AdS$_{d+1}$ black brane}

We now move on to discuss the inequalities of $\EW$ in the AdS black brane background in various dimensions. The corresponding metric in $d+1$ dimensions is
\begin{equation}
ds^2 = \frac{L_{\mathrm{AdS}}^2}{z^2} \left( -b(z)dt^2+ \frac{dz^2}{b(z)}+ dx^2 + d\vec{x}_{d-2}^2 \right) \ ,
\end{equation}
where $b(z)=1-\frac{z^d}{z_h^d}$ is the blackening factor and $z_h$ is the radius of the horizon.

\begin{figure}
\centering
    \includegraphics[width=0.8\textwidth]{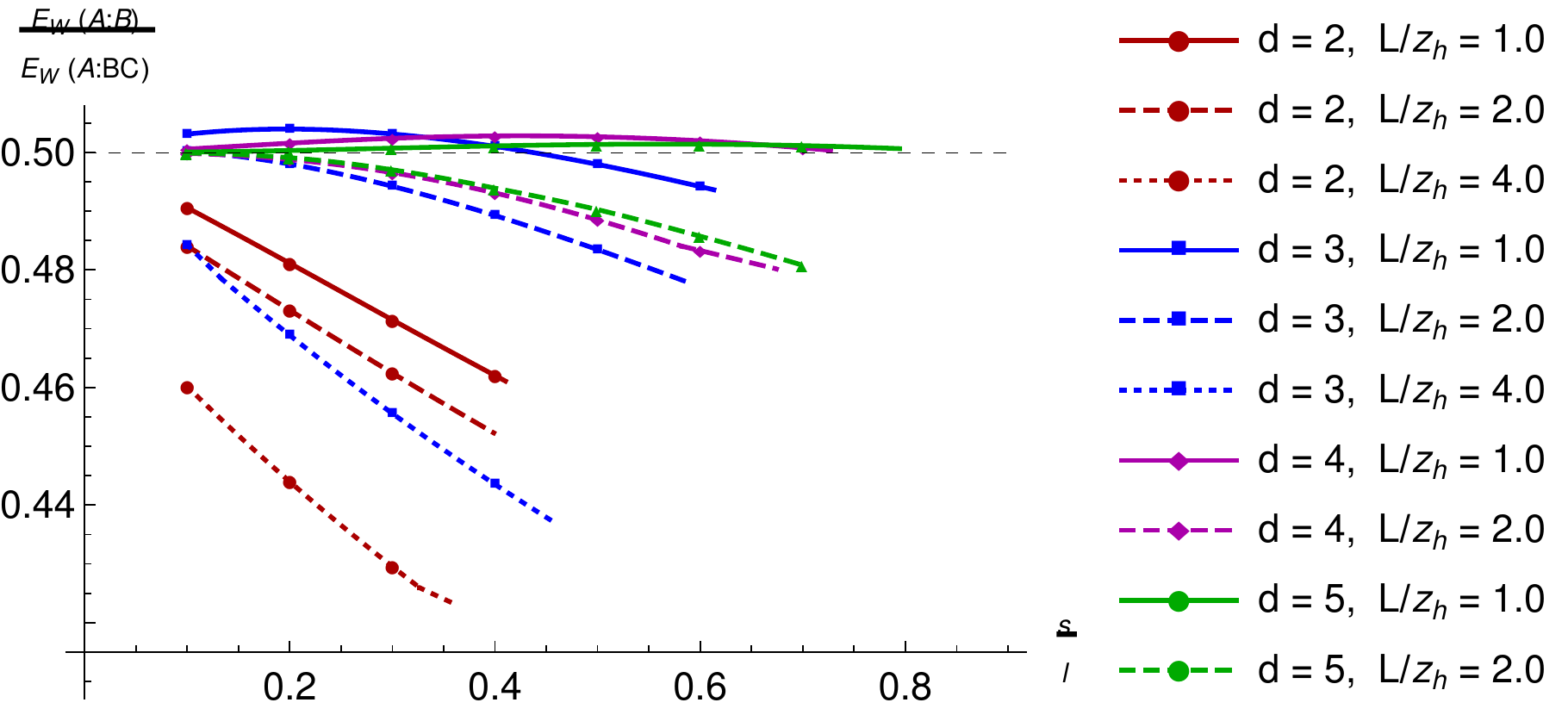}
    \caption{
The ratio of the wedges $\EW(A:B)/\EW(A:BC)$ in the $S_B$ phase as a function of $s/l$ for various $d$ and total widths $L$ in the $(d+1)$-dimensional AdS black brane background.
    }
    \label{fig:BB}
\end{figure}
\begin{figure}
\centering
    \includegraphics[width=0.6\textwidth]{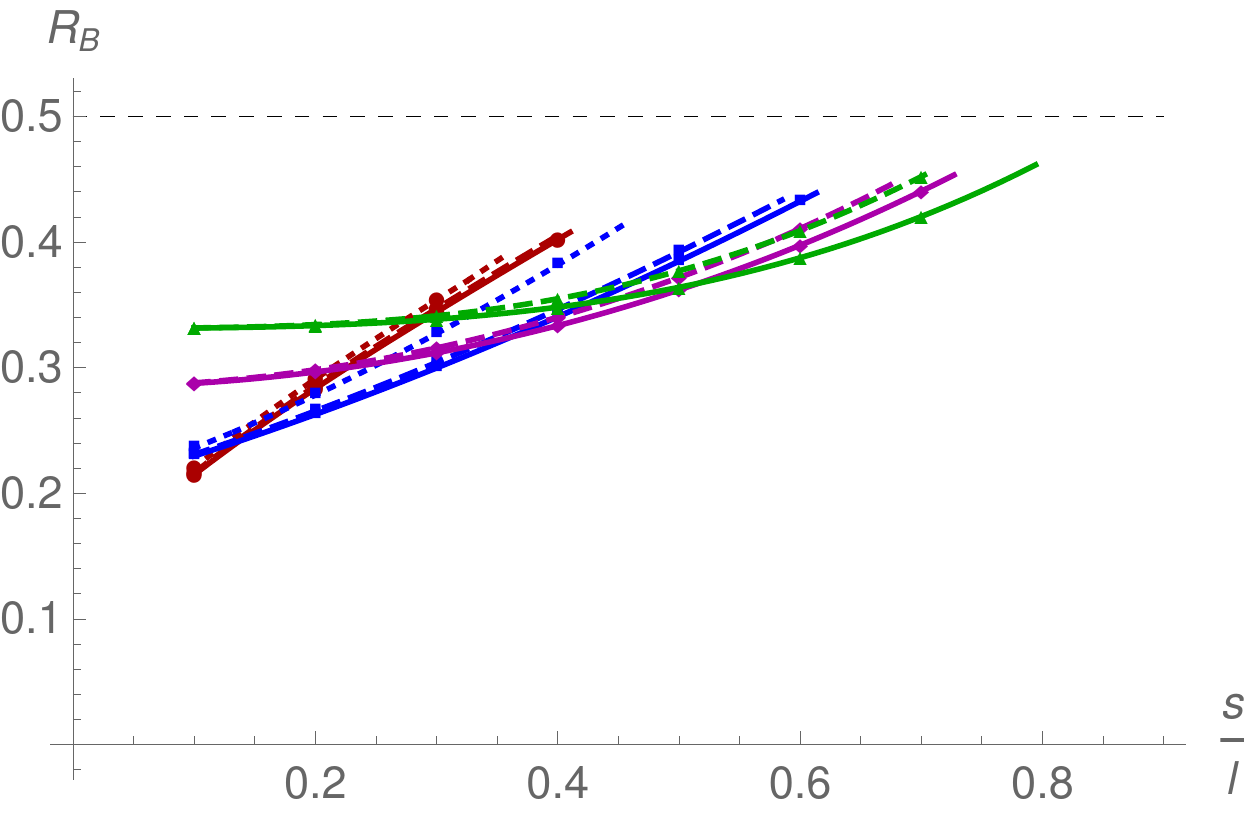}
    \caption{
Demonstration of weak monogamy in the $S_B$ phase as a function of $s/l$ for various $d$ and total widths $L$ in the $(d+1)$-dimensional AdS black brane background. Notation of curves as in Fig.~\protect\ref{fig:BB}. 
    }
    \label{fig:BB_ineq}
\end{figure}

We place the subsystems along the $x$-direction and again consider the symmetric configuration $l_1 = l_2 = l_3 = l$ and $s_1 = s_2 = s$ for simplicity. In the AdS black brane background, like in the pure AdS case, the wedge will exist only in the $S_B$ phase.  Our numerical results of the ratio $\EW(A:B)/\EW(A:BC)$ in the $S_B$ phase in various dimensions are shown in Fig.~\ref{fig:BB}. We observe that the ratio $\EW(A:B)/\EW(A:BC)$ is again less than one everywhere in the $S_B$ phase. Therefore, the inequality of Eq.~\eqref{TakaIneq} again holds true in the AdS black brane background. Moreover, the ratio $\EW(A:B)/\EW(A:BC)$ is also always less than $1/\sqrt{2}$. Accordingly, the squared version of monogamy inequality~\eqref{EWsq} and the corresponding lower bound~\eqref{ewlb} are satisfied everywhere in the $S_B$ phase.

However, the ratio $\EW(A:B)/\EW(A:BC)$ can be less than half only in some parts of the parameter space of $s$ and $l$, meaning that EWCS is neither monogamous nor polygamous. In particular, for large total width, \ie\ $L = 3l + 2s$, the ratio  $\EW(A:B)/\EW(A:BC)$ can be less than half even in the dominant $S_B$ phase, whereas for small $L$, the inequality can be preserved. RT surfaces tracing the black hole horizon can lead to the violation of the polygamy inequality. This means, just like in the pure AdS case, that the violation of the polygamy inequality depends on the dimension.

Notice also that some of the curves in Fig.~\ref{fig:BB} have kinks (\eg\ the curve with $d=2$, $L/z_h=4.0$) which are barely noticeable. The kinks reflect the competition between the two configurations in the middle row of Fig.~\ref{SAB}: for low $s/l$ the EWCS is determined by the standard single component configuration, whereas for large $s/l$ it is determined by the double component configuration. Actually, among the various classes of geometries considered in this article, the AdS black brane geometries is the class where the double component configuration contributes to our results.

Our results for the weak monogamy inequality $\EW(A:BC) + I(A:BC)/2\geq \EW(A:B) + \EW(A:C)$ in the $S_B$ phase for different values of $d$ and $L$ up to the bipartite critical point are shown in Fig.~\ref{fig:BB_ineq}. It shows that $R_B < 1/2$, hence lending support for weak monogamy for all $d$ and $L$ in the dominant $S_B$ phase.

\subsection{Non-conformal D$p$-brane backgrounds}\label{sec:Dpbackgrounds}

In this subsection we will analyze  the behavior of the inequalities as given in Sec.~\ref{sec:results} for non-conformal D$p$-brane backgrounds. The general form of the metric in the string frame is as follows 
\begin{equation}
 ds^2 = \left(\frac{r}{r_p}\right)^{\frac{7-p}{2}}\left(-dt^2 + \sum^{p}_{n=1}dx^2_n\right) 
+ \left(\frac{r_p}{r}\right)^{\frac{7-p}{2}}(dr^2
+ r^2 d\Omega^2_{8-p})\ ,
\end{equation}
and the dilaton is given as 
\begin{equation}
 e^{\phi} = \left(\frac{r_p}{r}\right)^{\frac{(7-p)(3-p)}{4}} \ .
\end{equation}
The radius of curvature $r_p$ is related to field theory variables $r^{7-p}_p \propto g^{2}_{\text{YM}}N$ with $g_{\text{YM}}$ being the Yang-Mills coupling and $N$ the rank of the gauge group \cite{Aharony:1999ti}. In these coordinates the boundary is at $r\to\infty$.
Here we are content with considering only the symmetric configurations, $l_1 = l_2 = l_3 = l$ and $s_1 = s_2 = s$. Also we will restrict to D$p$-brane backgrounds for $p<5$. The cases with $p>4$ are expected to correspond to non-local dual field theories, but one can nevertheless study their entanglement properties, see \cite{Barbon:2008ut,Kol:2014nqa} for more discussion.
The entanglement measures for $p<5$ have been considered in various works \cite{Bah:2007kcs,Pakman:2008ui,vanNiekerk:2011yi,Pang:2013lpa,Ben-Ami:2014gsa,Lala:2020lcp}.

\begin{figure}
\centering
    \includegraphics[width=0.8\textwidth]{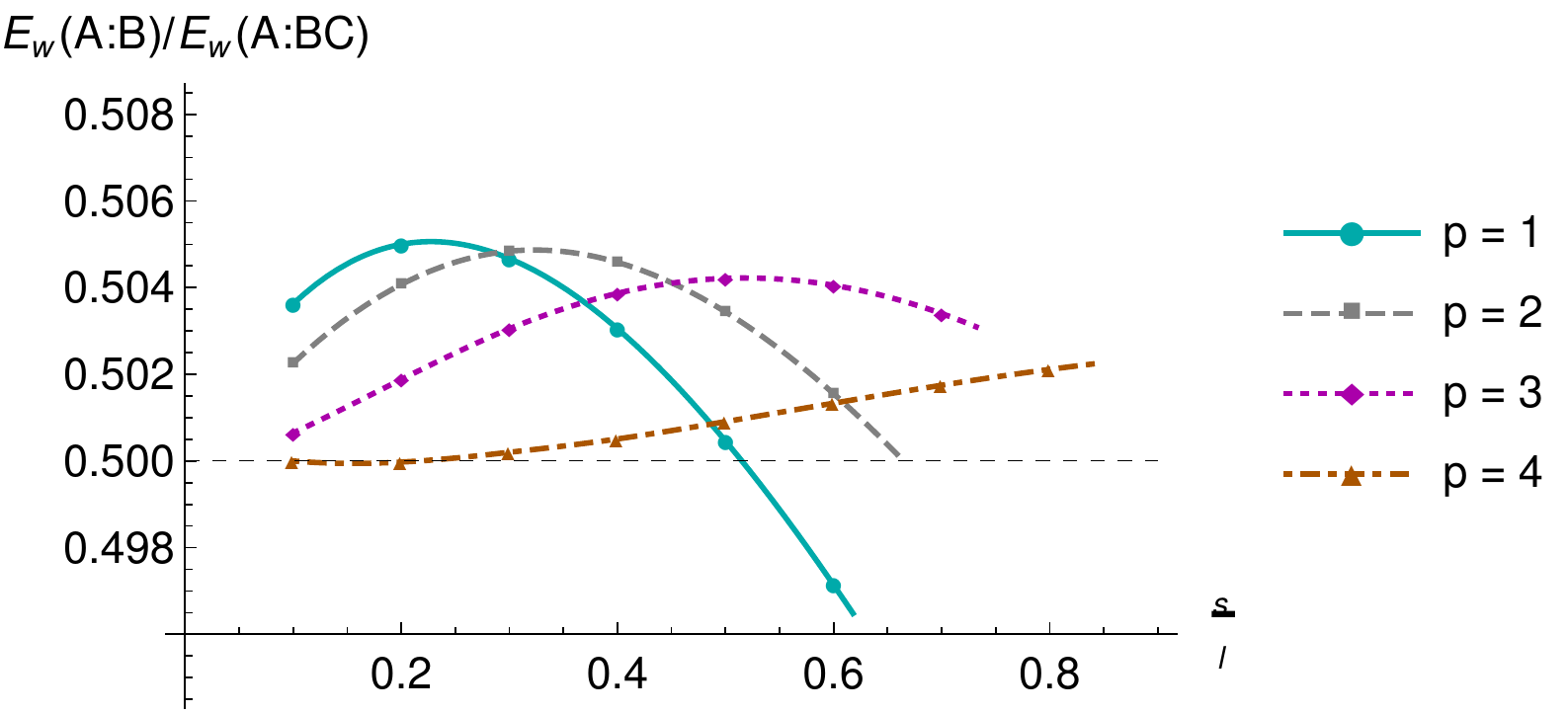}
    \caption{
The ratio of the wedges $\EW(A:B)/\EW(A:BC)$ in the $S_B$ phase for the D$p$-brane geometries as a function of $s/l$.     }
    \label{fig:Dp}
\end{figure}

We begin by analyzing the polygamy inequality (\ref{polygamycondition}). This is shown in Fig.~\ref{fig:Dp}. We find that for $p=1$ the inequality is violated as the ratio  $\EW(A:B)/\EW(A:BC)$ in the dominant $S_B$ phase goes below ${1}/{2}$. For $p=2$ and $p=3$ it is polygamous: one can check that the $p=2$ curve exits the $S_B$ phase before reaching $1/2$ as $s/l$ increases. We believe that the behavior is polygamous for $p=4$ also, but due to limited numerical accuracy, the data points at low $s/l$ are too close to $1/2$ to verify this. Recall that these results are for the configuration where $A$ is in the middle, see Fig.~\ref{SAB}. Similarly as for the AdS geometries, we find monogamous behavior for the configurations where $A$ is on the side for all $p<5$, so that   (\ref{polygamycondition}) is violated.

Furthermore, the Fig.~\ref{fig:Dp} shows that the ratio $\EW(A:B)/\EW(A:BC)$ in the $S_B$ phase is less than 1/$\sqrt{2}$. This means that the inequalities (\ref{TakaIneq}) and (\ref{EWsq}) are both satisfied. 

\begin{figure}
\centering
    \includegraphics[width=0.6\textwidth]{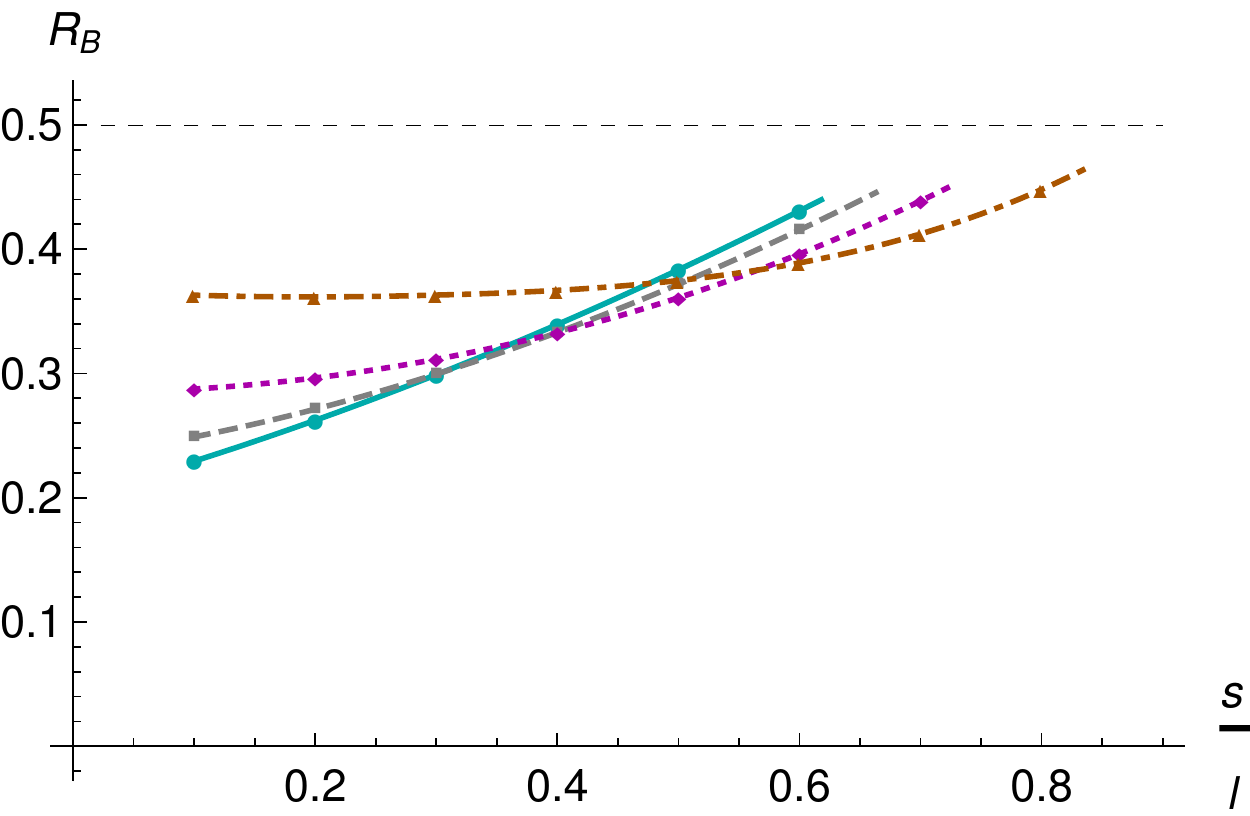}
    \caption{
Demonstration of weak monogamy in the $S_B$ phase for the D$p$-brane geometries as a function of $s/l$. Notation is similar to Fig.~\protect\ref{fig:Dp}. 
    }
    \label{fig:Dp_ineq}
\end{figure}

The weak monogamy as posited in (\ref{NewEWineqality}) is evinced in Fig.~\ref{fig:Dp_ineq}. The ratio $R_B$ again turns out to be less than 1/2, indicating weak monogamy.  

\section{Discussion and conclusions}\label{sec:discussion}

In this article, we studied the properties of the EWCS when we have a three disjoint slab system on the boundary. We focused on various gravitational bulk configurations and gave an analytic verification of various inequalities in the dominant $S_B$ phase  for the pure AdS$_{2+1}$ geometry. We also gave the numerical evidence for the same inequalities in the case of a generic pure AdS$_{d+1}$, AdS$_{d+1}$ soliton, AdS$_{d+1}$ black brane, and non-conformal D$p$-brane geometries. The results of our analysis are summarized below. 

\begin{enumerate}
\item  We showed that $\EW(A:BC)$ is not polygamous, \ie\ that the inequality (\ref{polygamycondition}) does not hold in general. 

\item  We gave evidence that the inequality (\ref{TakaIneq}) holds. 

\item We proposed that $\EW(A:BC)$ is weakly monogamous, \ie\ that the inequality (\ref{NewEWineqality}) holds in all the setups we considered in this work. 

\item We also proposed that the squared version of the inequality for $\EW(A:BC)$ holds. This is another monogamy property, measured via (\ref{EWsq}).

\item From our proposal and verification of the monogamy of $(\EW(A:BC))^2$, we provided a new lower bound for the linear version of $\EW(A:BC)$ by (\ref{ewlb}).
\end{enumerate}

The novel inequalities that we proposed may not, however, be the most stringent ones, but they nevertheless already alter conclusions in the literature as we have witnessed: The EWCS is at most weakly monogamous. A straightforward test-bed for the lower bound (\ref{EWsq}) for EWCS could be found in the AdS/BCFT context \cite{Li:2022mcv,Liu:2022ezb}. We note that an energy scale in the system played a pivotal role whenever expected behaviors for EWCS began faltering. All the cases studied in this paper, however, had the energy scale entering in a rather trivial way. It is therefore important to test the validity of the proposed weak monogamy in other more involved string theory setups, such as in those where there is a non-trivial renormalization group flow \cite{Bea:2013jxa,Balasubramanian:2018qqx,Jokela:2020wgs}.
Other interesting extensions would be to non-Lorentz invariant systems, \eg\ to those possessing intrinsic anisotropy \cite{Hoyos:2020zeg} or induced by an external control parameter \cite{Giataganas:2017koz,Arefeva:2018hyo,Gursoy:2018ydr,Chu:2019uoh,Gursoy:2020kjd,Jeong:2022zea,Caceres:2022hei,GonzalezLezcano:2022mcd,Jain:2022hxl,Bohra:2019ebj,Vasli:2022kfu}.

We believe that our program will eventually lead to an improved understanding of ``sharing of entanglement'' in strongly coupled systems.

\addcontentsline{toc}{section}{Acknowledgments}
\paragraph{Acknowledgments}
We would like to thank A.~P\"onni for discussions.  P.~J. and M.~J. have been supported by an appointment to the JRG Program at the APCTP through the Science and Technology Promotion Fund and Lottery Fund of the Korean Government. P.~J. and M.~J. have also been supported by the Korean Local Governments -- Gyeong\-sang\-buk-do Province and Pohang City -- and by the National Research Foundation of Korea (NRF) funded by the Korean government (MSIT) (grant number 2021R1A2C1010834). N.~J. is supported in part by the Academy of Finland grant. no 1322307. The work of S.~M. is supported by the Department of Science and Technology, Government of India under the Grant Agreement number IFA17-PH207 (INSPIRE Faculty Award).

\appendix

\section{Minimal surfaces and their areas}\label{app:surfaces}

In this Appendix we write down the basic formulas for minimal surfaces using a general metric, written as
\be
 ds^2 = -g_{tt}dt^2 + g_{zz}dz^2 + g_{xx}dx^2 + g_{\perp\perp}dx_{\perp}^2 + g_{cc}dx_c^2 \ ,
\ee
where we have singled out one field theory spatial direction $x$ from the rest that we denote by $x_{\perp}$. The direction $x_c$ is a spectator and could denote the possible compact direction or in the case of soliton geometry the pinching circle in which case care of counting dimensions is advised. 
Let us then consider slab regions on the boundary. We start with a single boundary region $A$ 
\be\label{eq:Aregime}
  A = \left\{ -\frac{l}{2}\leq x \leq \frac{l}{2},-\frac{L_2}{2}\leq x_2 \leq\frac{L_2}{2},\ldots,-\frac{L_{n}}{2}\leq x_n\leq \frac{L_{n}}{2}  \right\}  \ ,
\ee
where $n$ denotes the number of spatial directions.
We consider the directions $x_2,\ldots,x_n$ pertaining to $x_\perp$ to be periodic with periods $L_2,\ldots,L_{n}$, otherwise one ends up with partial differential equations. Eventually, we are interested in the limit $L_2,\ldots,L_n \to\infty$. We will consider the bulk surface $\Sigma$ ending on the boundary of $A$.  We will parametrize our embedding as $x=x(z)$. The induced metric on $\Sigma$ will be 
\be
 ds^2_{ind} = \left( g_{zz} + g_{xx}x'(z)^2 \right) dz^2 +g_{\perp\perp}dx_{\perp}^2 + g_{cc}dx_c^2 \ .
\ee
The equation of motion for the embedding reads
\be\label{xpara}
\frac{dx}{dz} = \frac{\sqrt{g_{zz}}}{\sqrt{g_{xx}}}\frac{1}{\sqrt{\frac{g_{xx}V_{\perp}^2}{g_{xx}V_{\perp}^2|_{z=z_{*}}}-1}} \ ,
\ee
where we have already fixed the integration constant such that the embedding is regular at the tip $z_*$, \ie\ demanding $x'(z_*)=0$. In this expression we have introduced $V_{\perp} = \sqrt{(g_{\perp\perp})^{n-1} g_{cc}}$. The area element for the minimal surface in the bulk becomes
\be\label{area}
d{\mathcal{A}} \propto V_{\perp} \sqrt{g_{zz} +g_{xx}x'(z)^2}\propto \frac{V_{\perp}\sqrt{g_{zz}}}{\sqrt{1-\frac{g_{xx}V_{\perp}^2|_{z=z_{*}}}{g_{xx}V_{\perp}^2}}}  \ .
\ee
Integrating this over all values of $z$ is (\ref{eq:hee2}), up to an overall coefficient that we do not need to keep track of in this work. 

In the applications of entanglement measures, we are interested in several boundary regions. For those, we simply shift in $x$ the corresponding regimes in (\ref{eq:Aregime}) and take linear combinations of individual entanglement entropies.  We will also use this general recipe to calculate the EWCS for various scenarios: for the configurations considered in this article, the surfaces needed for this are either straight or sections of the embedding considered in this Appendix.

\begin{figure}
\centering
\includegraphics[width=0.45\textwidth]{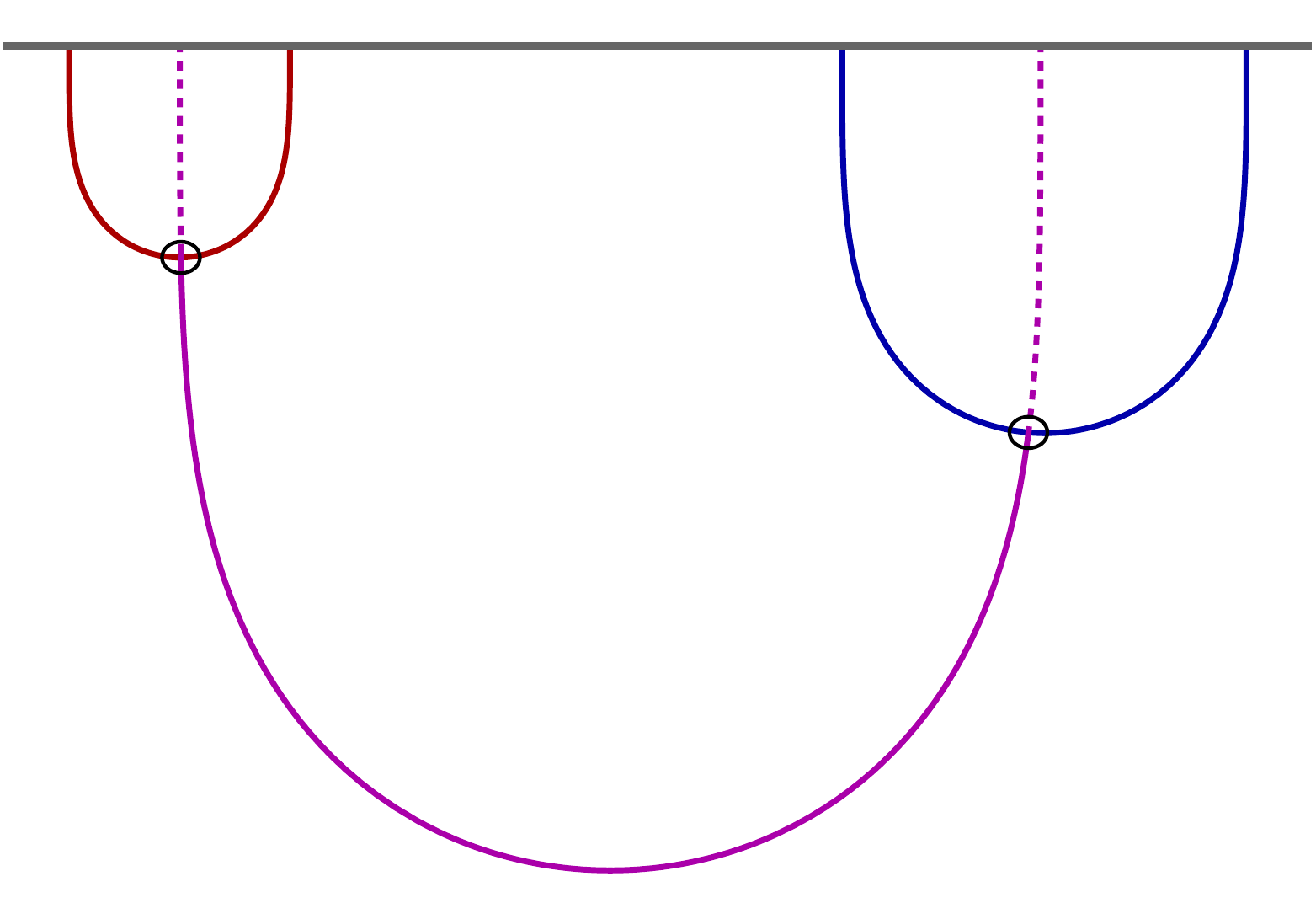}%
\hspace{10mm}\includegraphics[width=0.45\textwidth]{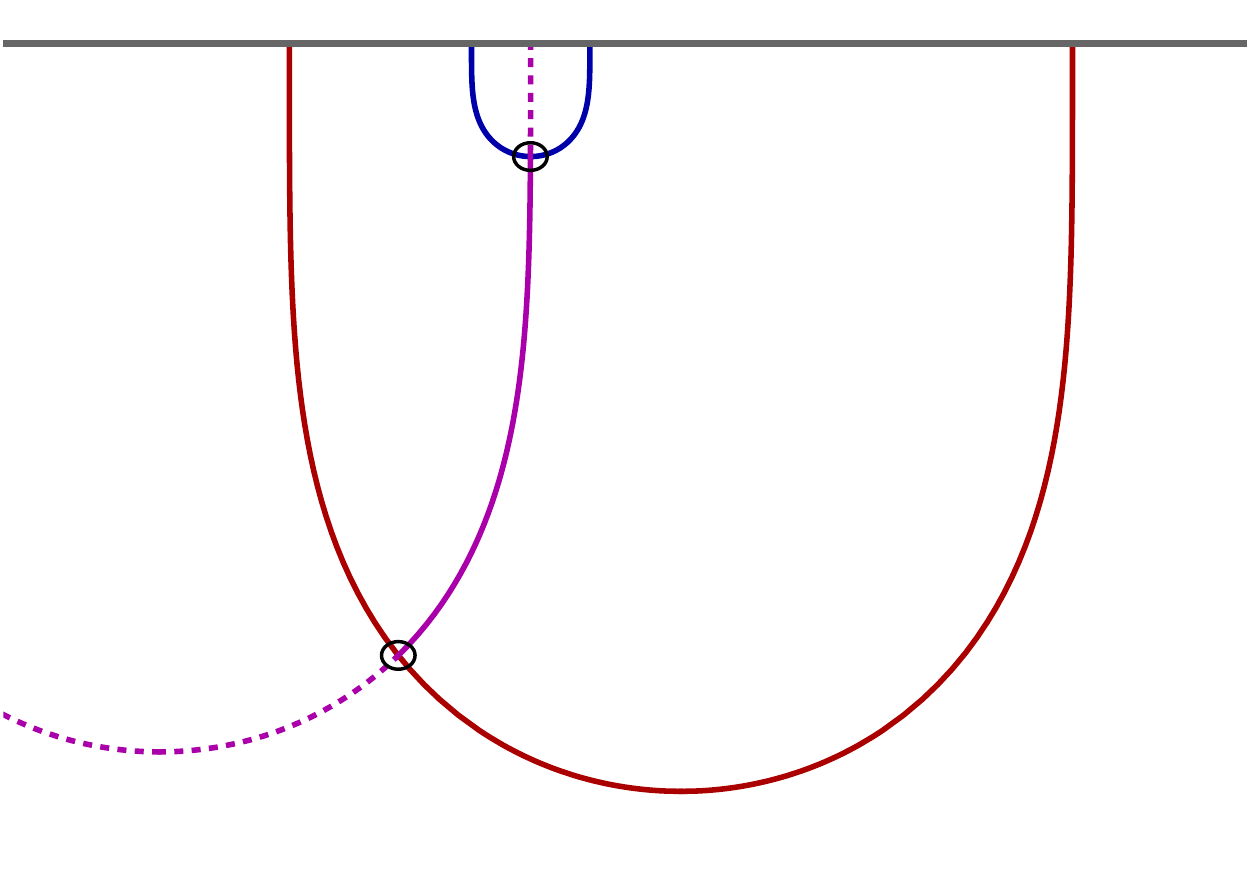}
\caption{The configurations appearing in the numerical minimization procedure for the EWCS. The red and blue curves are fixed (minimal) surfaces, parts of the boundary of the entanglement wedge. The magenta curve presents the cross section that needs to be minimized. }
\label{fig:wedgeexs} 
\end{figure}

\section{Numerical code for the entanglement wedge cross sections}\label{app:nummethod}

The EWCSs computed in this article are sometimes obtained by direct integration of closed form expressions over known intervals. However in the most interesting cases this does not happen: the location of the minimal surface, which gives (a contribution to) the EWCS, is not fixed by symmetry and needs to be found by numerically minimizing the area of the surface.

As it is clear from the definition~\eqref{EW}, the entanglement wedge is obtained by finding the surface (or a collection of surfaces) $\Sigma_{AB}$ that spans between different parts of $\Gamma_{AB}$. As one can show, the nontrivial minimization tasks boil down to two different tasks which are depicted in Fig.~\ref{fig:wedgeexs}: one needs to find the surface with minimal area spanning between the red and blue minimal surfaces, which are parts of $\Gamma_{AB}$. Typically $\Gamma_{AB}$ also contains other sections, which are irrelevant for the minimization, and not shown. The area obtained through the minimization (that of the solid magenta curve) is then the desired contribution to the entanglement wedge.

As the magenta curve is also a minimal surface, it can be extended to a complete surface reaching the AdS boundary (black horizontal line). The extensions are denoted as dashed curves. The end points of the extended curve must lie either in the areas enclosed by the red and blue curves (left configuration) or in the are enclosed by the blue curve and in the area outside the red curve (right configuration). This is because two minimal surfaces can intersect each other only once. 

The numerical computation is then done roughly as follows. A separate code is written to handle the two cases of Fig.~\ref{fig:wedgeexs} but they follow the same logic. First, one constructs the fixed minimal (blue and red) curves by numerical integration: as it is well know, the equation for the minimal surface reduces to a trivial first order differential equation that can be directly integrated. One then writes a function that does the same for the magenta curve but with end points that are otherwise arbitrary but constrained to the regions specified above. The function searches numerically for the intersection points with the blue and red curves (black circles in the figure), and returns the area of the solid part of the magenta curve, which is obtained through direct integration.

In the last step one then runs a standard minimization routine for the constructed function, \ie\ a routine that searches for the locations of the end points of the magenta curve that minimize the returned area. In some cases the convergence is only obtained when initial guesses for the end points are relatively close to the true solution. Therefore we scan over a set of initial guesses and choose the solution that leads to the smallest area.

\section{Critical points for pure AdS geometries}\label{app:critvals}


In order to calculate the critical point for $S_B$ to $S_A$ phase transition for any number of slabs and for any given $d>2$ in the case of a symmetric configuration for pure AdS, 
one needs to solve the following for positive and real root of $y$ which can be denoted as $y_c$ \cite{Ben-Ami:2014gsa}
\begin{eqnarray}\label{pureadscrit}
\frac{1}{(m+(m-1)y)^{d-2}} + \frac{m-1}{y^{d-2}} = m \ ,
\end{eqnarray}
where $m$ denotes the number of slabs on the boundary and $y=s/l$. For $d=3$ and $d=4$, a closed form expression for $y_c$ is known \cite{Ben-Ami:2014gsa}. 
For $d=2$, the critical points are obtained from
\begin{eqnarray}
y^{m-1}(m+(m-1)y)=1 \ ,
\end{eqnarray}
Using the above eqs. we have numerically calculated the bipartite and the tripartite critical point for various $d$ which are given in Table~\ref{bicrit}.

\bibliographystyle{JHEP}
\bibliography{biblio}

\providecommand{\href}[2]{#2}\begingroup\raggedright\begin{thebibliography}{10}

\bibitem{Ryu:2006ef}
S.~Ryu and T.~Takayanagi, \emph{{Aspects of Holographic Entanglement Entropy}},
  \href{https://doi.org/10.1088/1126-6708/2006/08/045}{\emph{JHEP} {\bfseries
  08} (2006) 045} [\href{https://arxiv.org/abs/hep-th/0605073}{{\ttfamily
  hep-th/0605073}}].

\bibitem{Ryu:2006bv}
S.~Ryu and T.~Takayanagi, \emph{{Holographic derivation of entanglement entropy
  from AdS/CFT}},
  \href{https://doi.org/10.1103/PhysRevLett.96.181602}{\emph{Phys. Rev. Lett.}
  {\bfseries 96} (2006) 181602}
  [\href{https://arxiv.org/abs/hep-th/0603001}{{\ttfamily hep-th/0603001}}].

\bibitem{Almheiri:2014lwa}
A.~Almheiri, X.~Dong and D.~Harlow, \emph{{Bulk Locality and Quantum Error
  Correction in AdS/CFT}},
  \href{https://doi.org/10.1007/JHEP04(2015)163}{\emph{JHEP} {\bfseries 04}
  (2015) 163} [\href{https://arxiv.org/abs/1411.7041}{{\ttfamily 1411.7041}}].

\bibitem{Dong:2016eik}
X.~Dong, D.~Harlow and A.~C. Wall, \emph{{Reconstruction of Bulk Operators
  within the Entanglement Wedge in Gauge-Gravity Duality}},
  \href{https://doi.org/10.1103/PhysRevLett.117.021601}{\emph{Phys. Rev. Lett.}
  {\bfseries 117} (2016) 021601}
  [\href{https://arxiv.org/abs/1601.05416}{{\ttfamily 1601.05416}}].

\bibitem{Bao:2015bfa}
N.~Bao, S.~Nezami, H.~Ooguri, B.~Stoica, J.~Sully and M.~Walter, \emph{{The
  Holographic Entropy Cone}},
  \href{https://doi.org/10.1007/JHEP09(2015)130}{\emph{JHEP} {\bfseries 09}
  (2015) 130} [\href{https://arxiv.org/abs/1505.07839}{{\ttfamily
  1505.07839}}].

\bibitem{Jokela:2020auu}
N.~Jokela and A.~P\"onni, \emph{{Towards precision holography}},
  \href{https://doi.org/10.1103/PhysRevD.103.026010}{\emph{Phys. Rev. D}
  {\bfseries 103} (2021) 026010}
  [\href{https://arxiv.org/abs/2007.00010}{{\ttfamily 2007.00010}}].

\bibitem{Headrick:2013zda}
M.~Headrick, \emph{{General properties of holographic entanglement entropy}},
  \href{https://doi.org/10.1007/JHEP03(2014)085}{\emph{JHEP} {\bfseries 03}
  (2014) 085} [\href{https://arxiv.org/abs/1312.6717}{{\ttfamily 1312.6717}}].

\bibitem{Hayden:2011ag}
P.~Hayden, M.~Headrick and A.~Maloney, \emph{{Holographic Mutual Information is
  Monogamous}}, \href{https://doi.org/10.1103/PhysRevD.87.046003}{\emph{Phys.
  Rev. D} {\bfseries 87} (2013) 046003}
  [\href{https://arxiv.org/abs/1107.2940}{{\ttfamily 1107.2940}}].

\bibitem{Headrick:2007km}
M.~Headrick and T.~Takayanagi, \emph{{A Holographic proof of the strong
  subadditivity of entanglement entropy}},
  \href{https://doi.org/10.1103/PhysRevD.76.106013}{\emph{Phys. Rev. D}
  {\bfseries 76} (2007) 106013}
  [\href{https://arxiv.org/abs/0704.3719}{{\ttfamily 0704.3719}}].

\bibitem{Czech:2012bh}
B.~Czech, J.~L. Karczmarek, F.~Nogueira and M.~Van~Raamsdonk, \emph{{The
  Gravity Dual of a Density Matrix}},
  \href{https://doi.org/10.1088/0264-9381/29/15/155009}{\emph{Class. Quant.
  Grav.} {\bfseries 29} (2012) 155009}
  [\href{https://arxiv.org/abs/1204.1330}{{\ttfamily 1204.1330}}].

\bibitem{Wall:2012uf}
A.~C. Wall, \emph{{Maximin Surfaces, and the Strong Subadditivity of the
  Covariant Holographic Entanglement Entropy}},
  \href{https://doi.org/10.1088/0264-9381/31/22/225007}{\emph{Class. Quant.
  Grav.} {\bfseries 31} (2014) 225007}
  [\href{https://arxiv.org/abs/1211.3494}{{\ttfamily 1211.3494}}].

\bibitem{Headrick:2014cta}
M.~Headrick, V.~E. Hubeny, A.~Lawrence and M.~Rangamani, \emph{{Causality \&
  holographic entanglement entropy}},
  \href{https://doi.org/10.1007/JHEP12(2014)162}{\emph{JHEP} {\bfseries 12}
  (2014) 162} [\href{https://arxiv.org/abs/1408.6300}{{\ttfamily 1408.6300}}].

\bibitem{Takayanagi:2017knl}
T.~Takayanagi and K.~Umemoto, \emph{{Entanglement of purification through
  holographic duality}},
  \href{https://doi.org/10.1038/s41567-018-0075-2}{\emph{Nature Phys.}
  {\bfseries 14} (2018) 573}
  [\href{https://arxiv.org/abs/1708.09393}{{\ttfamily 1708.09393}}].

\bibitem{Tamaoka:2018ned}
K.~Tamaoka, \emph{{Entanglement Wedge Cross Section from the Dual Density
  Matrix}}, \href{https://doi.org/10.1103/PhysRevLett.122.141601}{\emph{Phys.
  Rev. Lett.} {\bfseries 122} (2019) 141601}
  [\href{https://arxiv.org/abs/1809.09109}{{\ttfamily 1809.09109}}].

\bibitem{Dutta:2019gen}
S.~Dutta and T.~Faulkner, \emph{{A canonical purification for the entanglement
  wedge cross-section}},
  \href{https://doi.org/10.1007/JHEP03(2021)178}{\emph{JHEP} {\bfseries 03}
  (2021) 178} [\href{https://arxiv.org/abs/1905.00577}{{\ttfamily
  1905.00577}}].

\bibitem{Kudler-Flam:2018qjo}
J.~Kudler-Flam and S.~Ryu, \emph{{Entanglement negativity and minimal
  entanglement wedge cross sections in holographic theories}},
  \href{https://doi.org/10.1103/PhysRevD.99.106014}{\emph{Phys. Rev. D}
  {\bfseries 99} (2019) 106014}
  [\href{https://arxiv.org/abs/1808.00446}{{\ttfamily 1808.00446}}].

\bibitem{Kusuki:2019zsp}
Y.~Kusuki, J.~Kudler-Flam and S.~Ryu, \emph{{Derivation of holographic
  negativity in AdS$_3$/CFT$_2$}},
  \href{https://doi.org/10.1103/PhysRevLett.123.131603}{\emph{Phys. Rev. Lett.}
  {\bfseries 123} (2019) 131603}
  [\href{https://arxiv.org/abs/1907.07824}{{\ttfamily 1907.07824}}].

\bibitem{Bhattacharyya:2018sbw}
A.~Bhattacharyya, T.~Takayanagi and K.~Umemoto, \emph{{Entanglement of
  Purification in Free Scalar Field Theories}},
  \href{https://doi.org/10.1007/JHEP04(2018)132}{\emph{JHEP} {\bfseries 04}
  (2018) 132} [\href{https://arxiv.org/abs/1802.09545}{{\ttfamily
  1802.09545}}].

\bibitem{Hirai:2018jwy}
H.~Hirai, K.~Tamaoka and T.~Yokoya, \emph{{Towards Entanglement of Purification
  for Conformal Field Theories}},
  \href{https://doi.org/10.1093/ptep/pty063}{\emph{PTEP} {\bfseries 2018}
  (2018) 063B03} [\href{https://arxiv.org/abs/1803.10539}{{\ttfamily
  1803.10539}}].

\bibitem{Espindola:2018ozt}
R.~Esp\'\i{}ndola, A.~Guijosa and J.~F. Pedraza, \emph{{Entanglement Wedge
  Reconstruction and Entanglement of Purification}},
  \href{https://doi.org/10.1140/epjc/s10052-018-6140-2}{\emph{Eur. Phys. J. C}
  {\bfseries 78} (2018) 646}
  [\href{https://arxiv.org/abs/1804.05855}{{\ttfamily 1804.05855}}].

\bibitem{Bao:2018gck}
N.~Bao and I.~F. Halpern, \emph{{Conditional and Multipartite Entanglements of
  Purification and Holography}},
  \href{https://doi.org/10.1103/PhysRevD.99.046010}{\emph{Phys. Rev. D}
  {\bfseries 99} (2019) 046010}
  [\href{https://arxiv.org/abs/1805.00476}{{\ttfamily 1805.00476}}].

\bibitem{Umemoto:2018jpc}
K.~Umemoto and Y.~Zhou, \emph{{Entanglement of Purification for Multipartite
  States and its Holographic Dual}},
  \href{https://doi.org/10.1007/JHEP10(2018)152}{\emph{JHEP} {\bfseries 10}
  (2018) 152} [\href{https://arxiv.org/abs/1805.02625}{{\ttfamily
  1805.02625}}].

\bibitem{Yang:2018gfq}
R.-Q. Yang, C.-Y. Zhang and W.-M. Li, \emph{{Holographic entanglement of
  purification for thermofield double states and thermal quench}},
  \href{https://doi.org/10.1007/JHEP01(2019)114}{\emph{JHEP} {\bfseries 01}
  (2019) 114} [\href{https://arxiv.org/abs/1810.00420}{{\ttfamily
  1810.00420}}].

\bibitem{Caputa:2018xuf}
P.~Caputa, M.~Miyaji, T.~Takayanagi and K.~Umemoto, \emph{{Holographic
  Entanglement of Purification from Conformal Field Theories}},
  \href{https://doi.org/10.1103/PhysRevLett.122.111601}{\emph{Phys. Rev. Lett.}
  {\bfseries 122} (2019) 111601}
  [\href{https://arxiv.org/abs/1812.05268}{{\ttfamily 1812.05268}}].

\bibitem{Liu:2019qje}
P.~Liu, Y.~Ling, C.~Niu and J.-P. Wu, \emph{{Entanglement of Purification in
  Holographic Systems}},
  \href{https://doi.org/10.1007/JHEP09(2019)071}{\emph{JHEP} {\bfseries 09}
  (2019) 071} [\href{https://arxiv.org/abs/1902.02243}{{\ttfamily
  1902.02243}}].

\bibitem{Bhattacharyya:2019tsi}
A.~Bhattacharyya, A.~Jahn, T.~Takayanagi and K.~Umemoto, \emph{{Entanglement of
  Purification in Many Body Systems and Symmetry Breaking}},
  \href{https://doi.org/10.1103/PhysRevLett.122.201601}{\emph{Phys. Rev. Lett.}
  {\bfseries 122} (2019) 201601}
  [\href{https://arxiv.org/abs/1902.02369}{{\ttfamily 1902.02369}}].

\bibitem{BabaeiVelni:2019pkw}
K.~Babaei~Velni, M.~R. Mohammadi~Mozaffar and M.~H. Vahidinia, \emph{{Some
  Aspects of Entanglement Wedge Cross-Section}},
  \href{https://doi.org/10.1007/JHEP05(2019)200}{\emph{JHEP} {\bfseries 05}
  (2019) 200} [\href{https://arxiv.org/abs/1903.08490}{{\ttfamily
  1903.08490}}].

\bibitem{Jokela:2019ebz}
N.~Jokela and A.~P{\"o}nni, \emph{{Notes on entanglement wedge cross
  sections}}, \href{https://doi.org/10.1007/JHEP07(2019)087}{\emph{JHEP}
  {\bfseries 07} (2019) 087}
  [\href{https://arxiv.org/abs/1904.09582}{{\ttfamily 1904.09582}}].

\bibitem{Bao:2019wcf}
N.~Bao, A.~Chatwin-Davies, J.~Pollack and G.~N. Remmen, \emph{{Towards a Bit
  Threads Derivation of Holographic Entanglement of Purification}},
  \href{https://doi.org/10.1007/JHEP07(2019)152}{\emph{JHEP} {\bfseries 07}
  (2019) 152} [\href{https://arxiv.org/abs/1905.04317}{{\ttfamily
  1905.04317}}].

\bibitem{Harper:2019lff}
J.~Harper and M.~Headrick, \emph{{Bit threads and holographic entanglement of
  purification}}, \href{https://doi.org/10.1007/JHEP08(2019)101}{\emph{JHEP}
  {\bfseries 08} (2019) 101}
  [\href{https://arxiv.org/abs/1906.05970}{{\ttfamily 1906.05970}}].

\bibitem{Jeong:2019xdr}
H.-S. Jeong, K.-Y. Kim and M.~Nishida, \emph{{Reflected Entropy and
  Entanglement Wedge Cross Section with the First Order Correction}},
  \href{https://doi.org/10.1007/JHEP12(2019)170}{\emph{JHEP} {\bfseries 12}
  (2019) 170} [\href{https://arxiv.org/abs/1909.02806}{{\ttfamily
  1909.02806}}].

\bibitem{Bao:2019zqc}
N.~Bao and N.~Cheng, \emph{{Multipartite Reflected Entropy}},
  \href{https://doi.org/10.1007/JHEP10(2019)102}{\emph{JHEP} {\bfseries 10}
  (2019) 102} [\href{https://arxiv.org/abs/1909.03154}{{\ttfamily
  1909.03154}}].

\bibitem{Chu:2019etd}
J.~Chu, R.~Qi and Y.~Zhou, \emph{{Generalizations of Reflected Entropy and the
  Holographic Dual}},
  \href{https://doi.org/10.1007/JHEP03(2020)151}{\emph{JHEP} {\bfseries 03}
  (2020) 151} [\href{https://arxiv.org/abs/1909.10456}{{\ttfamily
  1909.10456}}].

\bibitem{Akers:2019gcv}
C.~Akers and P.~Rath, \emph{{Entanglement Wedge Cross Sections Require
  Tripartite Entanglement}},
  \href{https://doi.org/10.1007/JHEP04(2020)208}{\emph{JHEP} {\bfseries 04}
  (2020) 208} [\href{https://arxiv.org/abs/1911.07852}{{\ttfamily
  1911.07852}}].

\bibitem{Du:2019vwh}
D.-H. Du, F.-W. Shu and K.-X. Zhu, \emph{{Inequalities of Holographic
  Entanglement of Purification from Bit Threads}},
  \href{https://doi.org/10.1140/epjc/s10052-020-8283-1}{\emph{Eur. Phys. J. C}
  {\bfseries 80} (2020) 700}
  [\href{https://arxiv.org/abs/1912.00557}{{\ttfamily 1912.00557}}].

\bibitem{Kudler-Flam:2020url}
J.~Kudler-Flam, Y.~Kusuki and S.~Ryu, \emph{{Correlation measures and the
  entanglement wedge cross-section after quantum quenches in two-dimensional
  conformal field theories}},
  \href{https://doi.org/10.1007/JHEP04(2020)074}{\emph{JHEP} {\bfseries 04}
  (2020) 074} [\href{https://arxiv.org/abs/2001.05501}{{\ttfamily
  2001.05501}}].

\bibitem{Chandrasekaran:2020qtn}
V.~Chandrasekaran, M.~Miyaji and P.~Rath, \emph{{Including contributions from
  entanglement islands to the reflected entropy}},
  \href{https://doi.org/10.1103/PhysRevD.102.086009}{\emph{Phys. Rev. D}
  {\bfseries 102} (2020) 086009}
  [\href{https://arxiv.org/abs/2006.10754}{{\ttfamily 2006.10754}}].

\bibitem{Li:2020ceg}
T.~Li, J.~Chu and Y.~Zhou, \emph{{Reflected Entropy for an Evaporating Black
  Hole}}, \href{https://doi.org/10.1007/JHEP11(2020)155}{\emph{JHEP} {\bfseries
  11} (2020) 155} [\href{https://arxiv.org/abs/2006.10846}{{\ttfamily
  2006.10846}}].

\bibitem{Lala:2020lcp}
A.~Lala, \emph{{Entanglement measures for nonconformal D-branes}},
  \href{https://doi.org/10.1103/PhysRevD.102.126026}{\emph{Phys. Rev. D}
  {\bfseries 102} (2020) 126026}
  [\href{https://arxiv.org/abs/2008.06154}{{\ttfamily 2008.06154}}].

\bibitem{Jain:2020rbb}
P.~Jain and S.~Mahapatra, \emph{{Mixed state entanglement measures as probe for
  confinement}}, \href{https://doi.org/10.1103/PhysRevD.102.126022}{\emph{Phys.
  Rev. D} {\bfseries 102} (2020) 126022}
  [\href{https://arxiv.org/abs/2010.07702}{{\ttfamily 2010.07702}}].

\bibitem{Jokela:2020wgs}
N.~Jokela and J.~G. Subils, \emph{{Is entanglement a probe of confinement?}},
  \href{https://doi.org/10.1007/JHEP02(2021)147}{\emph{JHEP} {\bfseries 02}
  (2021) 147} [\href{https://arxiv.org/abs/2010.09392}{{\ttfamily
  2010.09392}}].

\bibitem{Wen:2021qgx}
Q.~Wen, \emph{{Balanced Partial Entanglement and the Entanglement Wedge Cross
  Section}}, \href{https://doi.org/10.1007/JHEP04(2021)301}{\emph{JHEP}
  {\bfseries 04} (2021) 301}
  [\href{https://arxiv.org/abs/2103.00415}{{\ttfamily 2103.00415}}].

\bibitem{Bao:2021vyq}
N.~Bao, A.~Chatwin-Davies and G.~N. Remmen, \emph{{Entanglement wedge cross
  section inequalities from replicated geometries}},
  \href{https://doi.org/10.1007/JHEP07(2021)113}{\emph{JHEP} {\bfseries 07}
  (2021) 113} [\href{https://arxiv.org/abs/2106.02640}{{\ttfamily
  2106.02640}}].

\bibitem{Basu:2021awn}
D.~Basu, A.~Chandra, V.~Raj and G.~Sengupta, \emph{{Entanglement wedge in flat
  holography and entanglement negativity}},
  \href{https://doi.org/10.21468/SciPostPhysCore.5.1.013}{\emph{SciPost Phys.
  Core} {\bfseries 5} (2022) 013}
  [\href{https://arxiv.org/abs/2106.14896}{{\ttfamily 2106.14896}}].

\bibitem{Hayden:2021gno}
P.~Hayden, O.~Parrikar and J.~Sorce, \emph{{The Markov gap for geometric
  reflected entropy}},
  \href{https://doi.org/10.1007/JHEP10(2021)047}{\emph{JHEP} {\bfseries 10}
  (2021) 047} [\href{https://arxiv.org/abs/2107.00009}{{\ttfamily
  2107.00009}}].

\bibitem{Akers:2021pvd}
C.~Akers, T.~Faulkner, S.~Lin and P.~Rath, \emph{{Reflected entropy in random
  tensor networks}}, \href{https://doi.org/10.1007/JHEP05(2022)162}{\emph{JHEP}
  {\bfseries 05} (2022) 162}
  [\href{https://arxiv.org/abs/2112.09122}{{\ttfamily 2112.09122}}].

\bibitem{Camargo:2022mme}
H.~A. Camargo, P.~Nandy, Q.~Wen and H.~Zhong, \emph{{Balanced partial
  entanglement and mixed state correlations}},
  \href{https://doi.org/10.21468/SciPostPhys.12.4.137}{\emph{SciPost Phys.}
  {\bfseries 12} (2022) 137}
  [\href{https://arxiv.org/abs/2201.13362}{{\ttfamily 2201.13362}}].

\bibitem{Jain:2022hxl}
P.~Jain, S.~S. Jena and S.~Mahapatra, \emph{{Holographic confining/deconfining
  gauge theories and entanglement measures with a magnetic field}},
  \href{https://arxiv.org/abs/2209.15355}{{\ttfamily 2209.15355}}.

\bibitem{Horodecki:2009zz}
R.~Horodecki, P.~Horodecki, M.~Horodecki and K.~Horodecki, \emph{{Quantum
  entanglement}}, \href{https://doi.org/10.1103/RevModPhys.81.865}{\emph{Rev.
  Mod. Phys.} {\bfseries 81} (2009) 865}
  [\href{https://arxiv.org/abs/quant-ph/0702225}{{\ttfamily
  quant-ph/0702225}}].

\bibitem{PhysRevA.61.052306}
V.~Coffman, J.~Kundu and W.~K. Wootters, \emph{Distributed entanglement},
  \href{https://doi.org/10.1103/PhysRevA.61.052306}{\emph{Phys. Rev. A}
  {\bfseries 61} (2000) 052306}.

\bibitem{PhysRevA.75.062308}
Y.-C. Ou and H.~Fan, \emph{Monogamy inequality in terms of negativity for
  three-qubit states},
  \href{https://doi.org/10.1103/PhysRevA.75.062308}{\emph{Phys. Rev. A}
  {\bfseries 75} (2007) 062308}.

\bibitem{PhysRevA.91.012339}
H.~He and G.~Vidal, \emph{Disentangling theorem and monogamy for entanglement
  negativity}, \href{https://doi.org/10.1103/PhysRevA.91.012339}{\emph{Phys.
  Rev. A} {\bfseries 91} (2015) 012339}.

\bibitem{Alishahiha:2014jxa}
M.~Alishahiha, M.~R. Mohammadi~Mozaffar and M.~R. Tanhayi, \emph{{On the Time
  Evolution of Holographic n-partite Information}},
  \href{https://doi.org/10.1007/JHEP09(2015)165}{\emph{JHEP} {\bfseries 09}
  (2015) 165} [\href{https://arxiv.org/abs/1406.7677}{{\ttfamily 1406.7677}}].

\bibitem{Mirabi:2016elb}
S.~Mirabi, M.~R. Tanhayi and R.~Vazirian, \emph{{On the Monogamy of Holographic
  $n$-partite Information}},
  \href{https://doi.org/10.1103/PhysRevD.93.104049}{\emph{Phys. Rev. D}
  {\bfseries 93} (2016) 104049}
  [\href{https://arxiv.org/abs/1603.00184}{{\ttfamily 1603.00184}}].

\bibitem{Nguyen:2017yqw}
P.~Nguyen, T.~Devakul, M.~G. Halbasch, M.~P. Zaletel and B.~Swingle,
  \emph{{Entanglement of purification: from spin chains to holography}},
  \href{https://doi.org/10.1007/JHEP01(2018)098}{\emph{JHEP} {\bfseries 01}
  (2018) 098} [\href{https://arxiv.org/abs/1709.07424}{{\ttfamily
  1709.07424}}].

\bibitem{PhysRevA.91.042323}
S.~Bagchi and A.~K. Pati, \emph{Monogamy, polygamy, and other properties of
  entanglement of purification},
  \href{https://doi.org/10.1103/PhysRevA.91.042323}{\emph{Phys. Rev. A}
  {\bfseries 91} (2015) 042323}.

\bibitem{PhysRevA.89.034303}
T.~R. de~Oliveira, M.~F. Cornelio and F.~F. Fanchini, \emph{Monogamy of
  entanglement of formation},
  \href{https://doi.org/10.1103/PhysRevA.89.034303}{\emph{Phys. Rev. A}
  {\bfseries 89} (2014) 034303}.

\bibitem{Calabrese:2009qy}
P.~Calabrese and J.~Cardy, \emph{{Entanglement entropy and conformal field
  theory}}, \href{https://doi.org/10.1088/1751-8113/42/50/504005}{\emph{J.
  Phys. A} {\bfseries 42} (2009) 504005}
  [\href{https://arxiv.org/abs/0905.4013}{{\ttfamily 0905.4013}}].

\bibitem{Hubeny:2007xt}
V.~E. Hubeny, M.~Rangamani and T.~Takayanagi, \emph{{A Covariant holographic
  entanglement entropy proposal}},
  \href{https://doi.org/10.1088/1126-6708/2007/07/062}{\emph{JHEP} {\bfseries
  07} (2007) 062} [\href{https://arxiv.org/abs/0705.0016}{{\ttfamily
  0705.0016}}].

\bibitem{Ben-Ami:2014gsa}
O.~Ben-Ami, D.~Carmi and J.~Sonnenschein, \emph{{Holographic Entanglement
  Entropy of Multiple Strips}},
  \href{https://doi.org/10.1007/JHEP11(2014)144}{\emph{JHEP} {\bfseries 11}
  (2014) 144} [\href{https://arxiv.org/abs/1409.6305}{{\ttfamily 1409.6305}}].

\bibitem{Bagchi_2015}
S.~Bagchi and A.~K. Pati, \emph{Monogamy, polygamy, and other properties of
  entanglement of purification},
  \href{https://doi.org/10.1103/physreva.91.042323}{\emph{Physical Review A}
  {\bfseries 91} (2015) }.

\bibitem{Aharony:2008ug}
O.~Aharony, O.~Bergman, D.~L. Jafferis and J.~Maldacena, \emph{{N=6
  superconformal Chern-Simons-matter theories, M2-branes and their gravity
  duals}}, \href{https://doi.org/10.1088/1126-6708/2008/10/091}{\emph{JHEP}
  {\bfseries 10} (2008) 091} [\href{https://arxiv.org/abs/0806.1218}{{\ttfamily
  0806.1218}}].

\bibitem{Maldacena:1997re}
J.~M. Maldacena, \emph{{The Large N limit of superconformal field theories and
  supergravity}}, \href{https://doi.org/10.1023/A:1026654312961}{\emph{Adv.
  Theor. Math. Phys.} {\bfseries 2} (1998) 231}
  [\href{https://arxiv.org/abs/hep-th/9711200}{{\ttfamily hep-th/9711200}}].

\bibitem{Balasubramanian:2018qqx}
V.~Balasubramanian, N.~Jokela, A.~P\"onni and A.~V. Ramallo, \emph{{Information
  flows in strongly coupled ABJM theory}},
  \href{https://doi.org/10.1007/JHEP01(2019)232}{\emph{JHEP} {\bfseries 01}
  (2019) 232} [\href{https://arxiv.org/abs/1811.09500}{{\ttfamily
  1811.09500}}].

\bibitem{Aharony:1999ti}
O.~Aharony, S.~S. Gubser, J.~M. Maldacena, H.~Ooguri and Y.~Oz, \emph{{Large N
  field theories, string theory and gravity}},
  \href{https://doi.org/10.1016/S0370-1573(99)00083-6}{\emph{Phys. Rept.}
  {\bfseries 323} (2000) 183}
  [\href{https://arxiv.org/abs/hep-th/9905111}{{\ttfamily hep-th/9905111}}].

\bibitem{Barbon:2008ut}
J.~L.~F. Barbon and C.~A. Fuertes, \emph{{Holographic entanglement entropy
  probes (non)locality}},
  \href{https://doi.org/10.1088/1126-6708/2008/04/096}{\emph{JHEP} {\bfseries
  04} (2008) 096} [\href{https://arxiv.org/abs/0803.1928}{{\ttfamily
  0803.1928}}].

\bibitem{Kol:2014nqa}
U.~Kol, C.~Nunez, D.~Schofield, J.~Sonnenschein and M.~Warschawski,
  \emph{{Confinement, Phase Transitions and non-Locality in the Entanglement
  Entropy}}, \href{https://doi.org/10.1007/JHEP06(2014)005}{\emph{JHEP}
  {\bfseries 06} (2014) 005} [\href{https://arxiv.org/abs/1403.2721}{{\ttfamily
  1403.2721}}].

\bibitem{Bah:2007kcs}
I.~Bah, A.~Faraggi, L.~A. Pando~Zayas and C.~A. Terrero-Escalante,
  \emph{{Holographic entanglement entropy and phase transitions at finite
  temperature}}, \href{https://doi.org/10.1142/S0217751X0904542X}{\emph{Int. J.
  Mod. Phys. A} {\bfseries 24} (2009) 2703}
  [\href{https://arxiv.org/abs/0710.5483}{{\ttfamily 0710.5483}}].

\bibitem{Pakman:2008ui}
A.~Pakman and A.~Parnachev, \emph{{Topological Entanglement Entropy and
  Holography}},
  \href{https://doi.org/10.1088/1126-6708/2008/07/097}{\emph{JHEP} {\bfseries
  07} (2008) 097} [\href{https://arxiv.org/abs/0805.1891}{{\ttfamily
  0805.1891}}].

\bibitem{vanNiekerk:2011yi}
A.~van Niekerk, \emph{{Entanglement Entropy in NonConformal Holographic
  Theories}},  \href{https://arxiv.org/abs/1108.2294}{{\ttfamily 1108.2294}}.

\bibitem{Pang:2013lpa}
D.-W. Pang, \emph{{Entanglement thermodynamics for nonconformal D-branes}},
  \href{https://doi.org/10.1103/PhysRevD.88.126001}{\emph{Phys. Rev. D}
  {\bfseries 88} (2013) 126001}
  [\href{https://arxiv.org/abs/1310.3676}{{\ttfamily 1310.3676}}].

\bibitem{Li:2022mcv}
P.~Li and Y.~Ling, \emph{{Inequalities for the holographic entanglement of
  purification in BCFT}},  \href{https://arxiv.org/abs/2206.13417}{{\ttfamily
  2206.13417}}.

\bibitem{Liu:2022ezb}
Y.~Liu, H.-D. Lyu and J.-K. Zhao, \emph{{Properties of gapped systems in
  AdS/BCFT}},  \href{https://arxiv.org/abs/2210.02802}{{\ttfamily 2210.02802}}.

\bibitem{Bea:2013jxa}
Y.~Bea, E.~Conde, N.~Jokela and A.~V. Ramallo, \emph{{Unquenched massive
  flavors and flows in Chern-Simons matter theories}},
  \href{https://doi.org/10.1007/JHEP12(2013)033}{\emph{JHEP} {\bfseries 12}
  (2013) 033} [\href{https://arxiv.org/abs/1309.4453}{{\ttfamily 1309.4453}}].

\bibitem{Hoyos:2020zeg}
C.~Hoyos, N.~Jokela, J.~M. Pen\'\i{}n and A.~V. Ramallo, \emph{{Holographic
  spontaneous anisotropy}},
  \href{https://doi.org/10.1007/JHEP04(2020)062}{\emph{JHEP} {\bfseries 04}
  (2020) 062} [\href{https://arxiv.org/abs/2001.08218}{{\ttfamily
  2001.08218}}].

\bibitem{Giataganas:2017koz}
D.~Giataganas, U.~G\"ursoy and J.~F. Pedraza, \emph{{Strongly-coupled
  anisotropic gauge theories and holography}},
  \href{https://doi.org/10.1103/PhysRevLett.121.121601}{\emph{Phys. Rev. Lett.}
  {\bfseries 121} (2018) 121601}
  [\href{https://arxiv.org/abs/1708.05691}{{\ttfamily 1708.05691}}].

\bibitem{Arefeva:2018hyo}
I.~Aref'eva and K.~Rannu, \emph{{Holographic Anisotropic Background with
  Confinement-Deconfinement Phase Transition}},
  \href{https://doi.org/10.1007/JHEP05(2018)206}{\emph{JHEP} {\bfseries 05}
  (2018) 206} [\href{https://arxiv.org/abs/1802.05652}{{\ttfamily
  1802.05652}}].

\bibitem{Gursoy:2018ydr}
U.~G\"ursoy, M.~J\"arvinen, G.~Nijs and J.~F. Pedraza, \emph{{Inverse
  Anisotropic Catalysis in Holographic QCD}},
  \href{https://doi.org/10.1007/JHEP04(2019)071}{\emph{JHEP} {\bfseries 04}
  (2019) 071} [\href{https://arxiv.org/abs/1811.11724}{{\ttfamily
  1811.11724}}].

\bibitem{Chu:2019uoh}
C.-S. Chu and D.~Giataganas, \emph{{$c$-Theorem for Anisotropic RG Flows from
  Holographic Entanglement Entropy}},
  \href{https://doi.org/10.1103/PhysRevD.101.046007}{\emph{Phys. Rev. D}
  {\bfseries 101} (2020) 046007}
  [\href{https://arxiv.org/abs/1906.09620}{{\ttfamily 1906.09620}}].

\bibitem{Gursoy:2020kjd}
U.~G\"ursoy, M.~J\"arvinen, G.~Nijs and J.~F. Pedraza, \emph{{On the interplay
  between magnetic field and anisotropy in holographic QCD}},
  \href{https://doi.org/10.1007/JHEP03(2021)180}{\emph{JHEP} {\bfseries 03}
  (2021) 180} [\href{https://arxiv.org/abs/2011.09474}{{\ttfamily
  2011.09474}}].

\bibitem{Jeong:2022zea}
H.-S. Jeong, K.-Y. Kim and Y.-W. Sun, \emph{{Holographic entanglement density
  for spontaneous symmetry breaking}},
  \href{https://doi.org/10.1007/JHEP06(2022)078}{\emph{JHEP} {\bfseries 06}
  (2022) 078} [\href{https://arxiv.org/abs/2203.07612}{{\ttfamily
  2203.07612}}].

\bibitem{Caceres:2022hei}
E.~Caceres and S.~Shashi, \emph{{Anisotropic Flows into Black Holes}},
  \href{https://arxiv.org/abs/2209.06818}{{\ttfamily 2209.06818}}.

\bibitem{GonzalezLezcano:2022mcd}
A.~Gonz\'alez~Lezcano, J.~Hong, J.~T. Liu, L.~A. Pando~Zayas and C.~F.
  Uhlemann, \emph{{c-functions in flows across dimensions}},
  \href{https://doi.org/10.1007/JHEP10(2022)083}{\emph{JHEP} {\bfseries 10}
  (2022) 083} [\href{https://arxiv.org/abs/2207.09360}{{\ttfamily
  2207.09360}}].

\bibitem{Bohra:2019ebj}
H.~Bohra, D.~Dudal, A.~Hajilou and S.~Mahapatra, \emph{{Anisotropic string
  tensions and inversely magnetic catalyzed deconfinement from a dynamical
  AdS/QCD model}},
  \href{https://doi.org/10.1016/j.physletb.2019.135184}{\emph{Phys. Lett. B}
  {\bfseries 801} (2020) 135184}
  [\href{https://arxiv.org/abs/1907.01852}{{\ttfamily 1907.01852}}].

\bibitem{Vasli:2022kfu}
M.~J. Vasli, M.~R. Mohammadi~Mozaffar, K.~Babaei~Velni and M.~Sahraei,
  \emph{{Holographic Study of Reflected Entropy in Anisotropic Theories}},
  \href{https://arxiv.org/abs/2207.14169}{{\ttfamily 2207.14169}}.

\end{thebibliography}\endgroup

\end{document}